\RequirePackage{fix-cm}
\documentclass[smallextended]{svjour3}       
\smartqed  
\usepackage{graphicx}
\usepackage[misc,geometry]{ifsym}
\usepackage{scrextend}
 \usepackage{geometry}
 \usepackage{float}
\usepackage{hyperref}
\usepackage{stmaryrd}
\usepackage{subcaption}
\usepackage[export]{adjustbox}
\usepackage[bottom]{footmisc}
\usepackage{bm}
\usepackage{amsmath,amsthm,amssymb}
\usepackage{textcomp}
\usepackage{psfrag}
\newgeometry{vmargin=25mm,hmargin={17mm,17mm}}
\renewcommand{\vec}[1]{\mathbf{#1}}
\newcommand{\be}{\begin{equation}}
\newcommand{\en}{\end{equation}}

\newcommand{\ep}{{\varepsilon}}
\newcommand{\paa}{\partial}

\journalname{Mechanics of Soft Materials}
\begin{document}
\title{Elasto-capillary circumferential buckling of soft tubes under axial loading: existence and competition with localised beading and periodic axial modes.}

\titlerunning{Mech Soft Mater}        

\author{Dominic Emery \and Yibin Fu}

\institute{Dominic Emery (\Letter) \and Yibin Fu \at
              School of Computing and Mathematics, Keele University, Staffordshire ST5 5BG, UK \\
\email{\{d.r.emery, y.fu\}@keele.ac.uk}}

\date{Received: 27 November 2020 / Accepted: date}

\maketitle

\begin{abstract}
We provide an extension to previous analysis of the localised beading instability of soft slender tubes under surface tension and axial stretching. The primary questions pondered here are: under what loading conditions, if any, can bifurcation into \textit{circumferential} buckling modes occur, and do such solutions dominate localisation and periodic axial modes? Three distinct boundary conditions are considered; in case 1 the tube's curved surfaces are traction-free and under surface tension, whilst in cases 2 and 3 the inner and outer surfaces (respectively) are fixed to prevent radial displacement and surface tension. A linear bifurcation analysis is conducted to determine numerically the existence of circumferential mode solutions. In case 1 we focus on the tensile stress regime given the preference of slender compressed tubes towards \textit{Euler buckling} over axial wrinkling. We show that tubes under several loading paths are highly sensitive to circumferential modes; in contrast, localised and periodic axial modes are absent, suggesting that the circumferential buckling is dominant by default. In case 2, circumferential mode solutions are associated with negative surface tension values and thus are physically implausible. Circumferential buckling solutions are shown to exist in case 3 for tensile \textit{and} compressive axial loads, and we demonstrate for multiple loading scenarios their dominance over localisation and periodic axial modes within specific parameter regimes.
\end{abstract}

\section{Introduction}
\label{intro}
Surface tension plays a dominant role in the finite deformation of non-linearly elastic materials below the \textit{elasto-capillary} length scale $\ell_{s}=\gamma/\mu$, where $\gamma$ is the surface tension and $\mu$ is the ground state shear modulus \cite{bico2018,lf2012,style}. Indeed, in extremely soft materials such as biological tissue or gels, or in solids with a sufficiently high surface area to volume ratio, this length can have an order of magnitude comparable to the microscale or even the milliscale. In such circumstances, it becomes imperative that surface tension is incorporated into the classical continuum framework. In recent years, elasto-capillary effects have been considered in many investigations of soft materials at finite strains. For instance, there have been exhaustive studies into surface instabilities in soft layers under the combined action of surface tension and uni-axial compression \cite{chen}, equi-biaxial strain \cite{ciarletta2014wrinkle} and growth \cite{alawiye2019lin}.

The beading instability of soft cylindrical tubes under axial loading and surface tension has received copious attention given its implication in the axonal degeneration caused by cytoskeletal trauma \cite{goriely2015,kilinc} and neurodegenerative disorders such as Alzheimer's and Parkinson's diseases \cite{datar2019}. Indeed, the theoretical nature of this bifurcation phenomenon is now extensively understood. Beading of a solid cylinder has been unanimously regarded as an infinite-wavelength instability through linear elasticity theory \cite{barriere1996,bc2003,mora2010}, non-linear elasticity theory \cite{ciarletta2012peristaltic,taffetani,xuan2016} and numerical approaches \cite{henann}. Explicitly, beading has been shown to be a phase separation phenomenon \cite{giudici2020,xuan2017}, and in \cite{FuST} it was demonstrated that, depending on the loading path that the cylinder is subjected to, several localised solutions can exist such as necking and bulging.

For a hollow tube, Finite Element Method simulations for the cases where the inner or outer curved surface is radially fixed were conducted in \cite{henann}. Where the outer radius tends to infinity, the special case of a cylindrical cavity inside an infinite solid is recovered. It was shown analytically in \cite{xuan2016} that a localised solution can exist for such a case. Recently, localisation in hollow tubes under two distinct boundary conditions was investigated in \cite{liuwang} under the framework of finite elasticity. Surprisingly, an analytical solution to the incremental eigenvalue problem was obtained in terms of modified Bessel functions. This was contrary to expectation given the investigations in \cite{HO1979}, where the governing equation for a tube under axial loading and internal pressure could only be solved numerically. This discrepancy was swiftly resolved in \cite{EmeryFu2020}, where we proved that the analysis in \cite{liuwang} is valid only where the primary deformation of the tube is homogeneous. Such a deformation may only occur in a hollow tube under the constraints of cases 2 and 3 with no axial stretching, a cylindrical cavity in an infinite solid or a solid cylinder, with the latter two having been analysed in \cite{xuan2016} and \cite{FuST}, respectively. In \cite{EmeryFu2020}, it was shown that localisation cannot occur in a tube where both curved surfaces are traction free and under surface tension. However, if one of the curved surfaces is radially fixed and free of surface tension, localisation is possible and preferred by the tube when the axial stress is tensile. Of course, this preference is strictly over bifurcation into periodic axial modes with non-zero wave number $k$. However, it may instead be the case that the tube develops finite circumferential buckling modes. Indeed, this has been extensively shown to be the case in single or multi-layered tubes under volumetric growth \cite{balbi2013,balbi2015,balbi2014,li2011}, as well as in a solid disk encapsulated by a growing layer under surface tension \cite{riccobelli2020}. The latter study gave a theoretical explanation for the genetic disorder \textit{lissencephaly} which is characterised by a reduction in brain sulci. It was shown that reduced stiffness of growing tissue (and thus increased surface tension) stabilised the layer against circumferential wrinkling modes. Here we investigate the potential circumferential buckling in soft tubes under axial load for all three cases alluded to in the abstract, whilst conducting an exhaustive analysis of the competition between possible solutions in tandem.

The remainder of this paper is structured as follows. After formulating the problem in section 2, we derive in section 3 the primary solution governing the axial loading of the tube under elasto-capillary effects, and recall the analytical bifurcation conditions for localisation given in \cite{EmeryFu2020}. In section 4 we conduct a linear bifurcation analysis from which we obtain a numerical bifurcation condition relating the load parameter (which we take to be the nominal axial stress $S_{zz}$ and the surface tension $\gamma$ separately) to the circumferential mode number $m$. From this relationship we deduce under which parameter regimes circumferential mode solutions can exist, and then analyse their competition with localisation and periodic axial modes where applicable. Concluding remarks are offered in section 5.

\section{Problem formulation}
\label{sec:2}

Consider a hyperelastic cylindrical tube whose reference and finitely deformed configurations are denoted by $\mathcal{B}_0$ and $\mathcal{B}_e$, respectively. A representative material particle in these configurations has the respective position vectors $\vec{X}=\vec{X}(R,\Theta,z)$ and $\vec{x}=\vec{x}(r,\vartheta,z)$ such that
\begin{align}
\vec{X}&=R\,\vec{E}_R + Z\,\vec{E}_Z,\,\,\,\,\,\,\,\,\vec{x}=r\,\vec{e}_r + z\,\vec{e}_z, \label{particlecoords}
\end{align}
where $\left(\vec{E}_R,\vec{E}_{\Theta},\vec{E}_Z\right)$ and $\left(\vec{e}_r,\vec{e}_{\vartheta},\vec{e}_z\right)$ are the corresponding orthonormal bases. The tube has inner and outer radii situated referentially at $R=A$ and $R=B$, and we denote by $a$ and $b$ the corresponding inner and outer radii in $\mathcal{B}_e$. The tube is assumed to have axial half-lengths $L$ and $\ell$ in $\mathcal{B}_0$ and $\mathcal{B}_e$, respectively.
\\\indent We consider a general deformation of the tube which is characterised by the following variable transformations
\begin{align}
r&=r\left(R,\Theta\right),\,\,\,\,\,\,\,\,\vartheta=\vartheta\left(R,\Theta\right),\,\,\,\,\,\,\,\,z=\lambda Z, \label{gendef}
\end{align}
where $\lambda=\ell/L$ is the principal axial stretch. The deformation gradient $\vec{F}=\paa \vec{x}/\paa \vec{X}$ then takes the following form
\begin{align}
\vec{F}&=\frac{\partial r}{\partial R}\,\vec{e}_r\otimes\vec{E}_R+\frac{1}{R}\frac{\partial r}{\partial \Theta}\,\vec{e}_r\otimes\vec{E}_{\Theta}+\frac{r}{R}\frac{\partial \vartheta}{\partial \Theta}\,\vec{e}_\vartheta\otimes\vec{E}_{\Theta}+r\frac{\partial \vartheta}{\partial R}\,\vec{e}_{\vartheta}\otimes\vec{E}_R+\lambda\,\vec{e}_z\otimes\vec{E}_Z. \label{Fgen}
\end{align}
The tube material is assumed to be incompressible, and so the following constraint of isochorism must be satisfied
\begin{align}
\text{det}\,\vec{F}&=1. \label{detF}
\end{align}
The constitutive behaviour of the tube is described by a strain energy function $W=W\left(I_B\right)$, where $I_B=\text{tr}\,\vec{B}$ is the first principal invariant of the left Cauchy-Green strain tensor $\vec{B}=\vec{F}\vec{F}^\top$ and the superscript $\top$ denotes transposition. We assume throughout this work that the tube material is neo-Hookean, for which the elastic strain energy density function is
\begin{align}
W(I_B)&=\frac{1}{2}\,\mu\left(I_B-3\right). \label{SEfunction}
\end{align}
For the remainder of this paper we scale all lengths by $B$, all stresses by $\mu$ and $\gamma$ by $\mu\,B$. Thus, we may set without loss of generality $\mu=1$ and $B=1$.
\textbf{\subsection{Stream function formulation}}
\label{sec:2p1}
As was originally proposed in \cite{ciarletta2011}, the problem can be elegantly reformulated in terms of a single mixed-coordinate stream function $\phi=\phi\left(R,\vartheta\right)$  which enforces the incompressibility constraint $(\ref{detF})$ exactly through the relations
\begin{align}
 r^2&=2\,\phi_{,\vartheta},\,\,\,\,\,\,\,\,\Theta=\frac{\lambda}{R}\,\phi_{,R}, \label{incphi}
\end{align}
where a comma denotes partial differentiation with respect to the implied coordinate. Thus, $\vec{F}$ as given by $(\ref{Fgen})$ can be expressed in terms of $\phi$ and its partial derivatives as such
\begin{align}
\vec{F}&=\frac{\left[\phi_{,R\vartheta}+\frac{\phi_{,\vartheta\vartheta}}{\phi_{,R\vartheta}}\left(\frac{\phi_{,R}}{R}-\phi_{,RR}\right)\right]}{\sqrt{2\,\phi_{,\vartheta}}}\,\vec{e}_r\otimes\vec{E}_R + \frac{\phi_{,\vartheta\vartheta}}{\lambda\sqrt{2\,\phi_{,\vartheta}}\,\phi_{,R\vartheta}}\,\vec{e}_r\otimes\vec{E}_{\Theta} + \frac{\sqrt{2\,\phi_{,\vartheta}}}{\lambda\,\phi_{,R\vartheta}}\,\vec{e}_{\vartheta}\otimes\vec{E}_{\Theta} \nonumber\\[1em]
&\,\,\,\,\,\,\,+\frac{\sqrt{2\,\phi_{,\vartheta}}}{\phi_{,R\vartheta}}\left[\frac{\phi_{,R}}{R}-\phi_{,RR}\right]\,\vec{e}_z\otimes\vec{E}_R + \lambda\,\vec{e}_z\otimes\vec{E}_Z. \label{Fphi}
\end{align}
It then follows that $I_B$ takes the form
\begin{align}
I_B=\frac{\left[\phi_{,R\vartheta}+\frac{\phi_{,\vartheta\vartheta}}{\phi_{,R\vartheta}}\left(\frac{\phi_{,R}}{R}-\phi_{,RR}\right)\right]^2}{2\,\phi_{,\vartheta}}+\frac{1}{2}\frac{\phi_{,\vartheta\vartheta}^2}{\lambda^2\,\phi_{,\vartheta}\,\phi_{R\vartheta}^2}+\frac{2\,\phi_{,\vartheta}}{\lambda^2\,\phi_{,R\vartheta}^2}+\frac{2\,\phi_{,\vartheta}}{\phi_{,R\vartheta}^2}\left[\frac{\phi_{,R}}{R}-\phi_{,RR}\right]^2+\lambda^2. \label{IBphi}
\end{align}

A variational approach is considered in deriving the bulk elastic equilibrium equations and the associated boundary conditions. The total potential energy $\mathcal{E}$ comprises of the bulk elastic energy $\mathcal{E}_b$ and the inner and outer surface energies $\mathcal{E}^{A}_{s}$ and $\mathcal{E}^B_s$ such that
\begin{align}
\mathcal{E}&=\mathcal{E}_b + \mathcal{E}^{A}_{s} + \mathcal{E}^{B}_s, \label{E}
\end{align}
where $\mathcal{E}_b$, $\mathcal{E}^{A}_{s}$ and $\mathcal{E}_s^B$ are given in terms of $\phi$ and its partial derivatives as follows
\begin{align}
\mathcal{E}_b&=\lambda\int^{2\,\pi}_{0}\int^{B}_{A}\,\phi_{,R\vartheta}\,W(I_B)\,dR\,d\vartheta,\,\,\,\,\,\,\,\,\mathcal{E}_{s}^{A,\,B}=\lambda\,\gamma\int^{2\,\pi}_{0}\,\left.\sqrt{2\phi_{,\vartheta}+\phi_{,\vartheta\vartheta}^{2}}\right\vert_{R=A,\,B}d\vartheta. \label{Ebphi}
\end{align}
Note that in $(\ref{Ebphi})$ $I_B$ is given by $(\ref{IBphi})$ and use has been made of the relation $d\Theta=\frac{\partial \Theta}{\partial \vartheta}d\vartheta=\frac{\lambda}{R}\,\phi_{,R\vartheta}\,d\vartheta$.
Equilibrium requires the vanishing of the first variation of $(\ref{Ebphi})_1$, i.e. $\delta\,\mathcal{E}_b=0$, which is equivalent to solving the Euler-Lagrange equations
\begin{align}
\left(\frac{\partial \mathcal{L}_b}{\partial \phi_{,iA}}\right)_{,iA}-\left(\frac{\partial \mathcal{L}_b}{\partial \phi_{,j}}\right)_{,j}&=0. \label{goveqn}
\end{align}
The standard summation convention is applied here, with $j=R$ or $\vartheta$ and $iA = RR$, $R\vartheta$ or $\vartheta\vartheta$. The bulk Lagrangian $\mathcal{L}_b$ is defined from $(\ref{Ebphi})_1$ as such
\begin{align}
\mathcal{L}_b&=\lambda\,\phi_{,R\vartheta}\,W(I_B). \label{bulklagrangian}
\end{align}
Now, in case 1 both curved boundaries $R=A$ and $R=B$ are traction-free and under surface tension, and these boundary conditions take the respective forms
\begin{align}
\left[\frac{\partial \mathcal{L}_b}{\partial \phi_{,R}}-\left(\frac{\partial \mathcal{L}_b}{\partial \phi_{,RR}}\right)_{,R}-\left(\frac{\partial \mathcal{L}_b}{\partial \phi_{,R\vartheta}}\right)_{,\vartheta}\right]_{R=A}-\left(\frac{\partial \mathcal{L}_s^{A}}{\partial \phi_{,\vartheta\vartheta}}\right)_{,\vartheta\vartheta}+\left(\frac{\partial \mathcal{L}_s^{A}}{\partial \phi_{,\vartheta}}\right)_{,\vartheta}&=0, \label{BC1A}
\end{align}
\begin{align}
\left[\frac{\partial \mathcal{L}_b}{\partial \phi_{,R}}-\left(\frac{\partial \mathcal{L}_b}{\partial \phi_{,RR}}\right)_{,R}-\left(\frac{\partial \mathcal{L}_b}{\partial \phi_{,R\vartheta}}\right)_{,\vartheta}\right]_{R=B}+\left(\frac{\partial \mathcal{L}_s^{B}}{\partial \phi_{,\vartheta\vartheta}}\right)_{,\vartheta\vartheta}-\left(\frac{\partial \mathcal{L}_s^{B}}{\partial \phi_{,\vartheta}}\right)_{,\vartheta}&=0, \label{BC3A}
\end{align}
where the inner and outer surface Lagrangian's $\mathcal{L}_s^A$ and $\mathcal{L}_s^B$ are given by
\begin{align}
\mathcal{L}_s^{A,\,B}&=\lambda\,\gamma\left.\sqrt{2\,\phi_{,\vartheta}+\phi_{,\vartheta\vartheta}^2}\right\vert_{R=A,\,B}. \label{surflagrangianA}
\end{align}
The opposite signs in the last two terms of $(\ref{BC1A})$ and $(\ref{BC3A})$ is due to the opposing curvatures of the inner and outer surfaces. In case 2 (resp. case 3), the inner (resp. outer) surface is constrained to prevent radial displacement, with the other curved boundary remaining traction free. Thus, we require that the incremental displacement in the $r$ direction, denoted $\delta r$, on this surface vanishes. 
For all three cases, zero shear on the inner and outer curved surfaces may be invoked by two further boundary conditions which are as follows
\begin{align}
\left.\frac{\partial \mathcal{L}_b}{\partial \phi_{,RR}}\right\vert_{R=A,\,B}&=0. \label{BC2}
\end{align}

\section{Primary deformation and conditions for localisation}
\label{sec:3}
We consider an axial loading of the tube, which is encapsulated by the following change of variables 
\begin{align}
r=r(R),\,\,\,\,\,\,\,\,\vartheta=\Theta,\,\,\,\,\,\,\,\,z=\lambda Z. \label{axdef}
\end{align}
A sub-class of $(\ref{gendef})$, the primary deformation $(\ref{axdef})$ is theoretically possible for all values of $\gamma$ and $\lambda$. The deformation gradient enforcing $(\ref{axdef})$ is
\begin{align}
\vec{F}&=\frac{\partial r}{\partial R}\,\vec{e}_r\otimes\vec{E}_R + \frac{r}{R}\,\vec{e}_{\vartheta}\otimes\vec{E}_{\Theta}+\lambda\,\vec{e}_z\otimes\vec{E}_Z. \label{Fax}
\end{align}
Upon substitution of $(\ref{Fax})$ into $(\ref{detF})$, incompressibility is found to be conditional on $r$ taking the following form  
\begin{align}
r(R)&=\sqrt{\lambda^{-1}(R^2-A^2)+a^2}. \label{rR}
\end{align}
It is then straightforward to deduce from $(\ref{rR})$ that the outer deformed radius $b=\sqrt{\lambda^{-1}\left(1-A^2\right)+a^2}$. Moreover, again with use of $(\ref{rR})$, the following primary solution for $\phi$, denoted $\phi_0$, which satisfies incompressibility exactly is deduced through integration of $(\ref{incphi})$
\begin{align}
\phi_{0}&=\frac{R^2\,\vartheta}{2\,\lambda} + \frac{1}{2}\left(a^2-\frac{A^2}{\lambda}\right)\vartheta. \label{phii0}
\end{align}
\textbf{\subsection{Case 1: Traction-free curved boundaries under surface tension}}
\label{sec:3p1}
In case 1, the inner radius $a$ in the primarily deformed state is an unknown quantity. We assume that the tube is under the combined action of surface tension and a nominal axial stress $S_{zz}$, with the latter being defined as the axial force per unit cross-sectional area in $\mathcal{B}_0$. Consequently, the total potential energy $\mathcal{E}$ as given in $(\ref{E})-(\ref{Ebphi})$ is modified as such
\begin{align}
\mathcal{E}=\mathcal{E}_b+\mathcal{E}_s^A+\mathcal{E}_s^B-\pi \left(1-A\right)^2\left(\lambda -1\right) S_{zz}. \label{Epr}
\end{align} 
For the primary deformation, $\mathcal{E}$ can be deduced by substituting $(\ref{phii0})$ into $(\ref{Epr})$. Equilibrium then requires that $\partial \mathcal{E}/\partial a=0$ and $\partial \mathcal{E}/\partial \lambda=0$. From the former, the following expression for $\gamma=\gamma\,(\lambda,\,a)$ can be obtained
\begin{align}
\gamma =\frac{\left(a^2\,\lambda-A^2\right)\left(b-a\right)}{2\,a\,b\,\lambda^2} + \frac{a\,b}{\lambda\left(a + b\right)}\,\log\left(\frac{A\,b}{a}\right). \label{gam}
\end{align}
From the latter, we deduce an expression for $S_{zz}=S_{zz}\left(\lambda,\,a\right)$ which is as follows
\begin{align}
S_{zz}=\frac{1}{2\,\lambda^2\left(1-A\right)^2} \Bigg[\Bigg.4\,a\,\gamma+\frac{a^2}{b^2}\left(A^2 -2\right)\left(\lambda^3 -1\right)+\frac{2\,\gamma\,\lambda}{b}\left(a^2+b^2\right)+2\,A^2\ln\left(\frac{a}{A\,b}\right)\Bigg.\Bigg]. \label{Nax}
\end{align}
For a localised solution to exist, we would anticipate that bifurcation into a mode characterised by zero axial wave number $k$ transpires \cite{Iooss,kirch}. It was shown in \cite{EmeryFu2020} that such a bifurcation can necessarily occur when the Jacobian $\mathcal{J}$ of the vector function $\left(\gamma,\,S_{zz}\right)$ vanishes. That is, the condition for localised bifurcation in case $1$ is
\begin{align}
\mathcal{J}\left(\gamma,\,S_{zz}\right)\equiv\frac{\partial \gamma}{\partial a}\frac{\partial S_{zz}}{\partial \lambda} - \frac{\partial \gamma}{\partial \lambda}\frac{\partial S_{zz}}{\partial a}=0. \label{bifconC1}
\end{align}
However, we then deduced that said condition is associated with negative surface tension values, which is physically implausible. Thus, localisation was deemed unattainable in case 1, and we showed that bifurcation into a periodic axial mode with non-zero wave number may necessarily take place instead provided that $S_{zz}$ is negative and of sufficient magnitude. The existence of circumferential mode solutions remains unresolved, however, and their materialisation is highly anticipated in lieu of localisation.

Notwithstanding, it was shown in \cite{goriely2008nonlinear} that slender cylindrical shells under axial \textit{compression} are of greater susceptibility to the \textit{Euler buckling} mode with axial wave number $k=\pi/\left(\lambda\,L\right)$ than periodic axial wrinkling. As such, for case 1 we focus in next section on the existence of and competition between periodic circumferential and axial modes specifically when the axial load is \textit{tensile}.

\textbf{\subsection{ Case 2: Radially fixed inner boundary free of surface tension}}
\label{sec:3p2}
In case 2, prevention of the radial displacement of the inner surface requires that $a=A$, and the absence of surface tension on this boundary enforces $\mathcal{E}_s^A=0$. Thus, the primary deformation is determined absolutely here by $\lambda$, and $\phi_0$ and $b$ become
\begin{align}
\phi_{0}&=\frac{R^2\,\vartheta}{2\,\lambda} + \frac{A^2}{2}\left(1-\frac{1}{\lambda}\right)\vartheta,\,\,\,\,\,\,\,\,b=\sqrt{\lambda^{-1}\left(1-A^2\right)+A^2}. \label{phii0C2}
\end{align}
Hence, to ensure equilibrium we need only satisfy the single equation $\partial \mathcal{E}/\partial \lambda=0$. From said equation we obtain the following expression for $S_{zz}=S_{zz}\left(\lambda\right)$
\begin{align}
S_{zz}&=\frac{1}{\pi^2\left(1-A\right)^2}\Bigg[\Bigg.\frac{\left(1-\lambda\right)}{2\,\lambda^2}\left(\frac{A^4}{b^2}+\left(2\,\lambda +1\right)\left(A^2 -\lambda\right)-\lambda -2\right)+\frac{\gamma}{b}\left(A^2 +b^2\right)-\frac{A^2}{\lambda^2}\ln b\Bigg.\Bigg]. \label{axialforcecase2}
\end{align}
The bifurcation condition for localisation was then shown in \cite{EmeryFu2020} to be $\partial S_{zz}/\partial \lambda =0$, from which the following expression for the critical surface tension can be obtained
\begin{align}
\gamma_{cr}&=\frac{1}{b\,\lambda^{2}\left(A^2-1\right)}\left[\frac{4 \left(A^3 (\lambda -1)+A\right)^2 \ln b}{A^2-1}-A^4\,\xi_{1}(\lambda) -A^2\,\xi_{2}(\lambda)-2 \left(\lambda ^3+2\right)\right],
\end{align}
where $\xi_{1}(\lambda)=2 \lambda ^5-4 \lambda ^4+2 \lambda ^3+2 \lambda ^2-3 \lambda +1$ and $\xi_{2}(\lambda)=4 \lambda ^4-4 \lambda ^3+7 \lambda -5$.

\textbf{\subsection{ Case 3: Radially fixed outer boundary free of surface tension}}
\label{sec:3p3}
In case 3, the radial fixing of the outer boundary requires that $b=B$. With use of $(\ref{rR})$, this invokes the following expression for $a$
\begin{align}
a=\sqrt{\lambda^{-1}\left(A^2 -1\right)+1}.
\end{align} 
Moreover, the absence of surface tension on said boundary requires that $\mathcal{E}_s^B=0$. The primary solution $\phi_0$ therefore is reduced to the following
\begin{align}
\phi_{0}&=\frac{R^2\,\vartheta}{2\,\lambda} + \frac{1}{2}\left(1-\frac{1}{\lambda}\right)\vartheta. \label{phii0C3}
\end{align}
Then, equilibrium again requires only that we set $\partial \mathcal{E}/\partial \lambda = 0$, and the following expression for $S_{zz}=S_{zz}(\lambda)$ is subsequently obtained
\begin{align}
S_{zz}&=\frac{1}{2\,\pi\,\lambda^2\left(1-A\right)^2}\Bigg[\Bigg.\frac{2\,\gamma\,\lambda^2}{a}\left(1+a^2\right)+\left(\lambda -1\right)\left(\frac{1}{a^2}+\lambda +1\right)-2\,A^2\left(\lambda^3 -1\right)+2\ln\left(\frac{a}{A}\right)\Bigg.\Bigg]. \label{axialforce}
\end{align}
Once more, the bifurcation condition for localisation is that $\partial S_{zz}/\partial \lambda=0$, from which we obtained in \cite{EmeryFu2020} the following expression for the critical surface tension $\gamma_{cr}$
\begin{align}
\gamma_{cr}&=\frac{a}{\lambda^2\left(A^2-1\right)^2}\left[\left(2-2\,A^2\right)\,\lambda^4+\lambda+\lambda^2-4\,A^2\,\lambda-\lambda\,\ln\left(\frac{a^4}{A^4}\right)+\frac{2+\lambda -\lambda^2}{a^4}\right]. \label{gamcr}
\end{align}
With the outer surface fixed to prevent radial displacement, Euler buckling is circumvented in favour of conventional periodic wrinkling modes. Indeed this statement is intuitive, and in \cite{liu2018axial} it was shown for the case of purely mechanical axial compression that axial or circumferential modes with wave number $n \pi/\left(\lambda L\right)$ ($n\neq 1$) will be favoured by the tube in case 3. Thus, unlike in case 1, we may extend our linear bifurcation analysis in the next section for case 3 to compressive axial loads; this will facilitate an exhaustive investigation into the competition between localisation and periodic axial and circumferential modes when elasto-capillary effects are taken into consideration.\\

\section{Linear bifurcation analysis}
\label{sec:4}

To begin, we look for a solution of the form
\begin{align}
\phi=\phi_{0}+\ep\,f(R)\,e^{i m \vartheta}, \label{pertsol}
\end{align}
where $\ep$ is a small parameter, $f$ is a scalar function of $R$ and $m$ is the circumferential mode number. 
On substituting $(\ref{pertsol})$ into $(\ref{goveqn})$, we obtain a fourth order ordinary differential equation (ODE) for $f$, which may be re-written as the following system of first order linear ODEs;
\begin{align}
\frac{d \bm{f}}{d R}&=\textsf{A}(R)\bm{f},\,\,\,\,\,\,\,\,
\textsf{A}=\begin{bmatrix}
0 & 1 & 0 & 0 \\
0 & 0 & 1 & 0 \\
0 & 0 & 0 & 1 \\
a_{41} & a_{42} & a_{43} & a_{44}
\end{bmatrix}, \label{goveq}
\end{align}
where $\bm{f}=\left[\,f,\,f',\,f'',\,f'''\,\right]^{\top}$ and the variable components of $\textsf{A}$ are given as follows
\begin{equation}
\begin{aligned}
a_{41}(R)&=\frac{m^2}{\eta^3}\left(\frac{2 \eta ^2}{R^2}+2 R^2-\eta\,m^2\right),\,\,\,\,\,\,\,\,a_{42}(R)=\frac{m^2 R}{\eta ^2}+\frac{2 \left(m^2-2\right)}{\eta  R}-\frac{2\,m^2 R^3}{\eta ^3}-\frac{3 \left(m^2-1\right)}{R^3},\\[1em]
a_{43}(R)&=\frac{4}{\eta }+\frac{m^2 R^2}{\eta ^2}-\frac{3-m^2}{R^2},\,\,\,\,\,\,\,\,a_{44}(R)=\frac{2}{R}-\frac{4 R}{\eta},
\end{aligned}
\end{equation}
with $\eta\equiv\eta(R)=R^2-A^2+a^2\,\lambda$.

On substituting $(\ref{pertsol})$ into $(\ref{BC1A})-(\ref{surflagrangianA})$ and $(\ref{BC2})$, we find that the boundary conditions on $R=A$ and $R=B$ in case 1 may be expressed respectively as the following matrix equations
\begin{equation}
\begin{aligned}
\textsf{B}_1(A,\xi)\,\bm{f}&=\bm{0}, \\[1em]
\textsf{B}_2(B,\xi)\,\bm{f}&=\bm{0},
\end{aligned}\,\,\,\,\,\,\,\,\,\,\,\,\,\,\,\,\text{where}\,\,\,\,\,\,\,\,\,\,\,\,\,\,\,\,\begin{cases}
\textsf{B}_1(R,\xi)=\begin{bmatrix}
b_{11} & -1/R & 1 & 0 \\
b_{21}^{+} & b_{22} & b_{23} & 1
\end{bmatrix}, \\[2em]
\textsf{B}_2(R,\xi)=\begin{bmatrix}
b_{11} & -1/R & 1 & 0 \\
b_{21}^{-} & b_{22} & b_{23} & 1
\end{bmatrix},
\end{cases}
\end{equation}
with
\begin{equation}
\begin{aligned}
b_{11}(R,\,\xi)&=\left(\frac{m R}{\eta}\right)^2,\,\,\,\,\,\,\,\,b_{22}(R,\,\xi)=\frac{1-m^2}{R^2}-\frac{2 m^2 R^2}{\eta^2}-\frac{2}{\eta},\,\,\,\,\,\,\,\,b_{23}(R,\,\xi)=-\frac{1}{2}\,a_{44},\\[1em]
b_{21}^{\pm}(R,\,\xi)&=\frac{m^2}{\eta^3\,R}\left(\left(\eta +R^2\right)^2\pm\gamma\,\left(\eta\,\lambda\right)^{1/2} R^2 \left(m^2 \left(a^2 \lambda +R^2\right)-A^2 m^2-\lambda \right)\right).
\end{aligned}
\end{equation}
Note that $\xi$ is a dummy variable introduced for presentational purposes to represent the load parameter, for which there can be several choices. For cases 2 and 3, on substituting $(\ref{pertsol})$ into $(\ref{incphi})_1$, we deduce that satisfying $\delta\,r=0$ on the surfaces $R=A$ and $B$ requires we enforce the corresponding constraints $f(A)=0$ and $f(B)=0$ in place of traction-free conditions. Indeed, the matrices $\textsf{B}_1$ and $\textsf{B}_2$ can then be modified appropriately. Now, the linear system $\textsf{B}_1(A,\xi)\,\bm{f}=\bm{0}$ can be shown to have two independent solutions, say $\bm{f}_0^{(1)}$ and $\bm{f}_0^{(2)}$. For instance, in case 1 we have
\begin{equation}
\begin{aligned}
\bm{f}_0^{(1)}&=\Big[\,1,\,0,\,-b_{11},\,b_{23}\,b_{11}-b^{+}_{21}\,\Big]^{\top}_{R=A},\,\,\,\,\,\,\,\,\bm{f}_0^{(2)}&=\Big[\,0,\,1,\,-1/A,\,b_{23}/A-b_{22}\,\Big]^{\top}_{R=A}. \label{IDa}
\end{aligned}
\end{equation}
We may then integrate forward $(\ref{goveq})$ from $R=A$ to $R=B$, using $(\ref{IDa})$ or equivalent as initial data for $\bm{f}$ at $R=A$. Two linearly independent solutions for $\bm{f}$, say $\bm{f}_{1}$ and $\bm{f}_2$ are obtained, and thus a general solution for $\bm{f}$ takes the form
\begin{align}
\bm{f}=c_1\,\bm{f}_1 + c_2\,\bm{f}_2 = \textsf{M}(R,\,\xi)\,\bm{c}, \label{gensol}
\end{align}
where $\bm{c}=\left[\,c_1,\,c_2\,\right]^{\top}$ is an arbitrary constant vector and $\textsf{M}(R,\,\xi)=\left[\,\bm{f}_1,\,\bm{f}_2\,\right]$. By its construction, $(\ref{gensol})$ satisfies the boundary conditions at $R=A$, and it thus remains only to satisfy the corresponding conditions on $R=B$. On substituting $(\ref{gensol})$ into $\textsf{B}_2(B,\xi)\bm{f}=\bm{0}$, we obtain $\textsf{B}_2\,\textsf{M}\,(B,\,\xi)\,\bm{c}=\bm{0}$. Then, since $\bm{c}$ is arbitrary, the existence of a non-trivial solution to the eigenvalue problem is conditional on satisfying 
\begin{align}
\text{det}\,\big[\,\textsf{B}_2\,\textsf{M}\,(B,\,\xi)\,\big]=0. \label{bfcn}
\end{align}
Indeed, $(\ref{bfcn})$ represents a numerical bifurcation condition which must be satisfied by $\lambda$, $\gamma$ and $m$. The bifurcation points are obtained by iterating on the load parameter until $(\ref{bfcn})$ is satisfied. We may take either $S_{zz}$ or $\gamma$ as the load parameter, and we indeed consider both of these cases in the following analysis. In the former, the tube is firstly subjected to a fixed surface tension $\gamma$ and zero axial load. Then, we vary the nominal axial stress $S_{zz}$ monotonically from zero, with $S_{zz}<0$ corresponding to a compressive stress and $S_{zz}>0$ being a tensile stress. In the latter, we first fix $\lambda$ and then gradually increase $\gamma$ from zero.

The primary aim of this linear bifurcation analysis is to produce a relationship between the load parameter $\xi$ and the circumferential mode number $m$. This enables us to deduce firstly whether bifurcation into a finite circumferential mode is possible. Then, if it is, we can ascertain the critical load $\xi_{cr}$ and circumferential mode number $m_{cr}$. The latter must be an integer greater than or equal to two in order to ensure the periodicity of the solution. By definition, $m_{cr}$ characterises the circumferential buckling mode \textit{preferred} by the tube. As we vary the load parameter from zero, $m_{cr}$ is the first integer value of $m$ encountered on the bifurcation curve in the $\left(m,\,\xi\right)$ plane. Then, $\xi_{cr}$ is the value of $\xi$ at this point and the critical load at which bifurcation into the preferred circumferential mode occurs.

We note briefly that all presented results were obtained through the previously outlined \textit{determinant method}. However, for verification purposes, we also performed an equivalent analysis using the \textit{compound matrix method} \cite{ng1985}, and excellent agreement was observed between both approaches. The algebraic manipulations and numerical solution procedures presented in this paper were implemented in \textbf{Mathematica} \cite{wo2019}. Throughout the ensuing analysis, we will compare the regimes of existence of circumferential and axial modes. For a detailed account of the linear bifurcation analysis for the latter case, the reader is referred to \cite{EmeryFu2020}.
\textbf{\subsection{ Case 1: Traction-free curved boundaries under surface tension}}
We start by applying a fixed stretch $\lambda\geq 1$ and take $\gamma$ as the load parameter. In  Figure $4$ \textbf{(a)} of \cite{EmeryFu2020}, it was shown that neither localised nor periodic axial mode solutions can exist under such a loading path. Thus, if circumferential mode solutions do occur, we can conclude that they will be dominant by default.

\begin{figure}[h!]
\centering
\begin{subfigure}[t]{0.465\textwidth}
\includegraphics[width=\linewidth, valign=t]{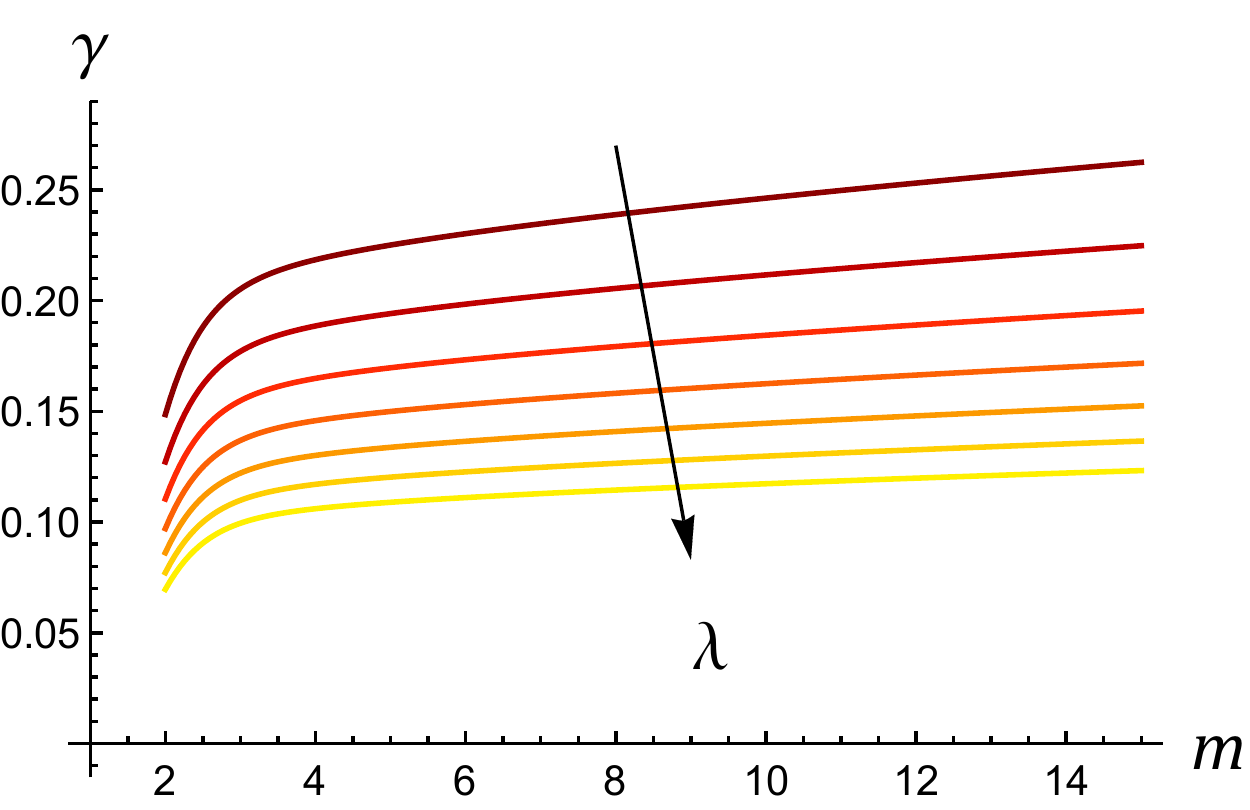}
\subcaption*{\textbf{(a)}}
\end{subfigure}\hfill
\begin{subfigure}[t]{0.46\textwidth}
\includegraphics[width=\linewidth, valign=t]{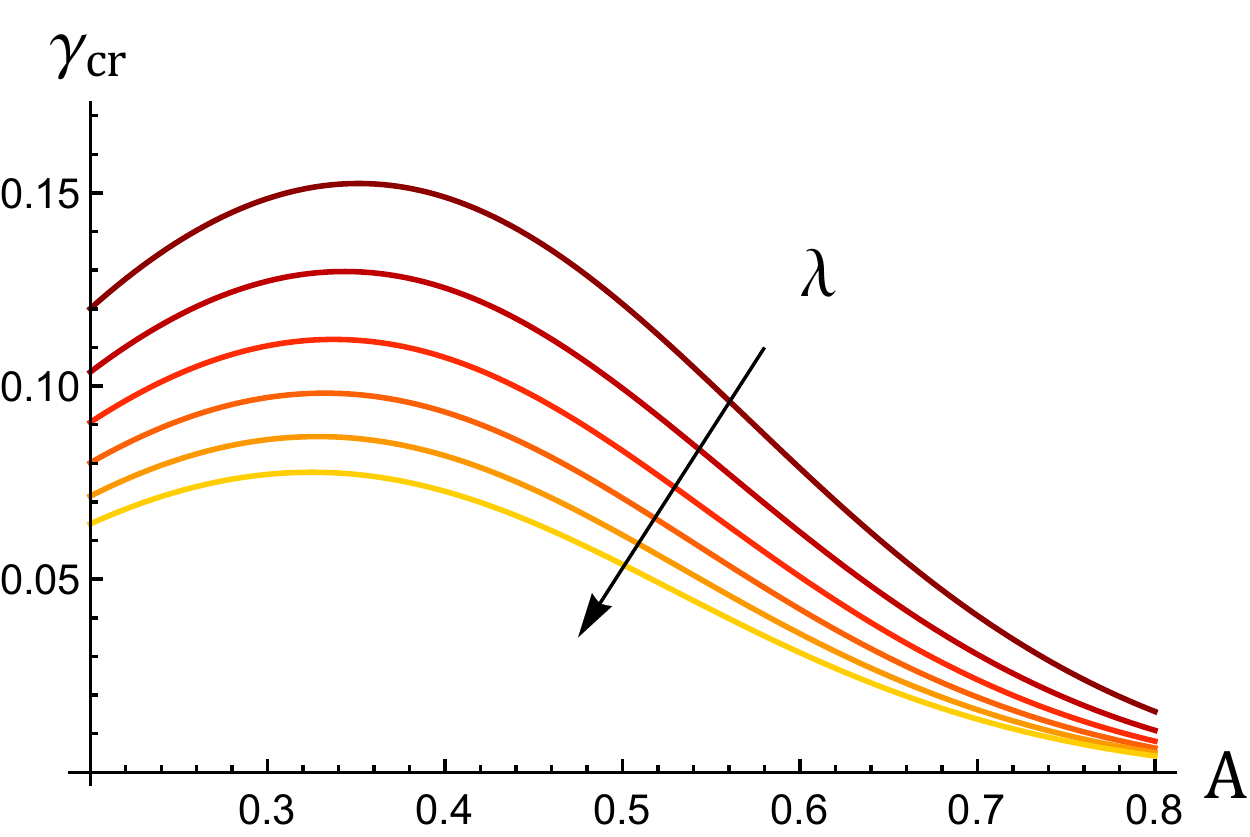}
\subcaption*{\textbf{(b)}}
\end{subfigure}
\caption{\textbf{(a)} The variation of $\gamma$ with respect to $m$ for $A=0.3$. The axial stretch $\lambda$ is increased from $1$ to $1.6$ in increments of $0.1$. \textbf{(b)} Conditions for bifurcation into the critical circumferential mode $m_{cr}=2$. Plotted is $\gamma_{cr}$ against $A$ with $\lambda$ increased from $1$ to $1.5$ in increments of $0.1$. Arrows indicate the direction of parameter growth.}
\label{fig1}
\end{figure}

\begin{figure}[h!]
\centering
\begin{subfigure}[t]{0.465\textwidth}
\includegraphics[width=\linewidth, valign=t]{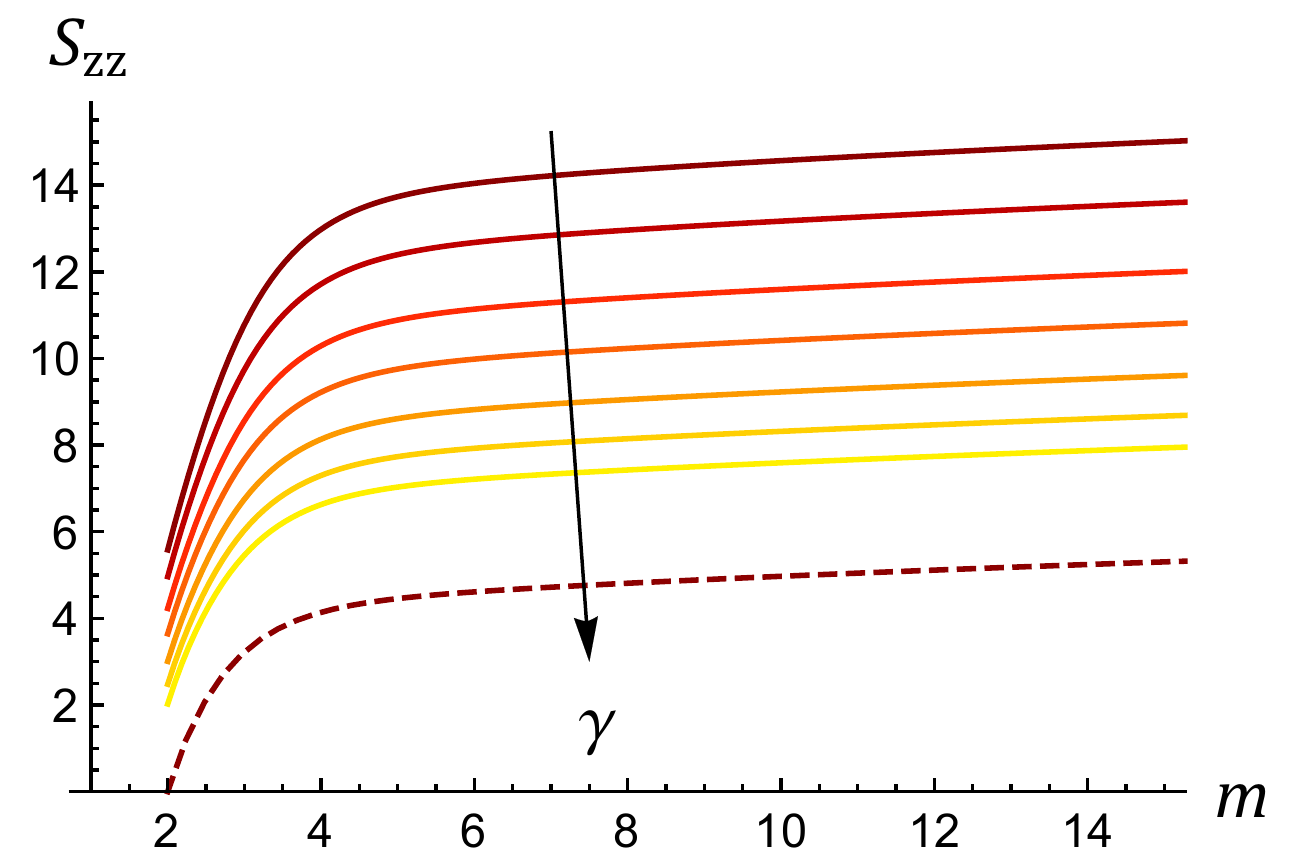}
\subcaption*{\textbf{(a)}}
\end{subfigure}\hfill
\begin{subfigure}[t]{0.45\textwidth}
\includegraphics[width=\linewidth, valign=t]{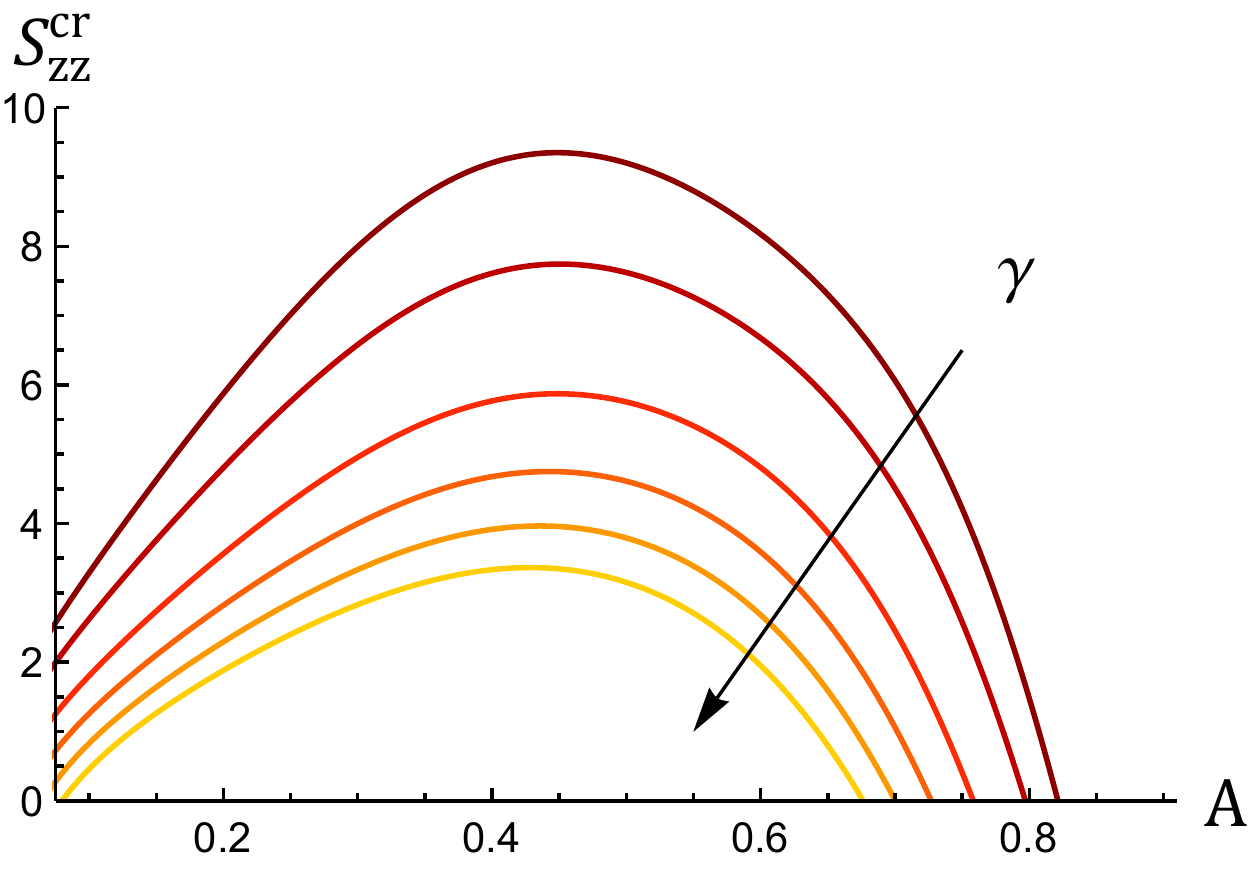}
\subcaption*{\textbf{(b)}}
\end{subfigure}
\caption{\textbf{(a)} The variation of $S_{zz}$ with respect to the circumferential mode number $m$ for $A=0.5$ and $\gamma=0.03,\,0.035,\,0.425,\,0.05,\,0.06,\,0.07\,\,\text{and}\,\,0.08$. The lowest-most dashed curve corresponds to $\gamma=0.142375$. At this point the tube becomes highly unstable with $S_{zz}^{cr}=0$. \textbf{(b)} Conditions for bifurcation into the elliptic mode $m_{cr}=2$. Plotted is $S_{zz}^{cr} $ against $A$ for $\gamma=0.01,\,0.015,\,0.02,\,0.03,\,0.04\,\,\text{and}\,\,0.05$. Arrows indicate the direction of parameter growth.}
\label{fig2}
\end{figure}

\begin{figure}[h!]
\centering
\begin{subfigure}[t]{0.47\textwidth}
\includegraphics[width=\linewidth, valign=t]{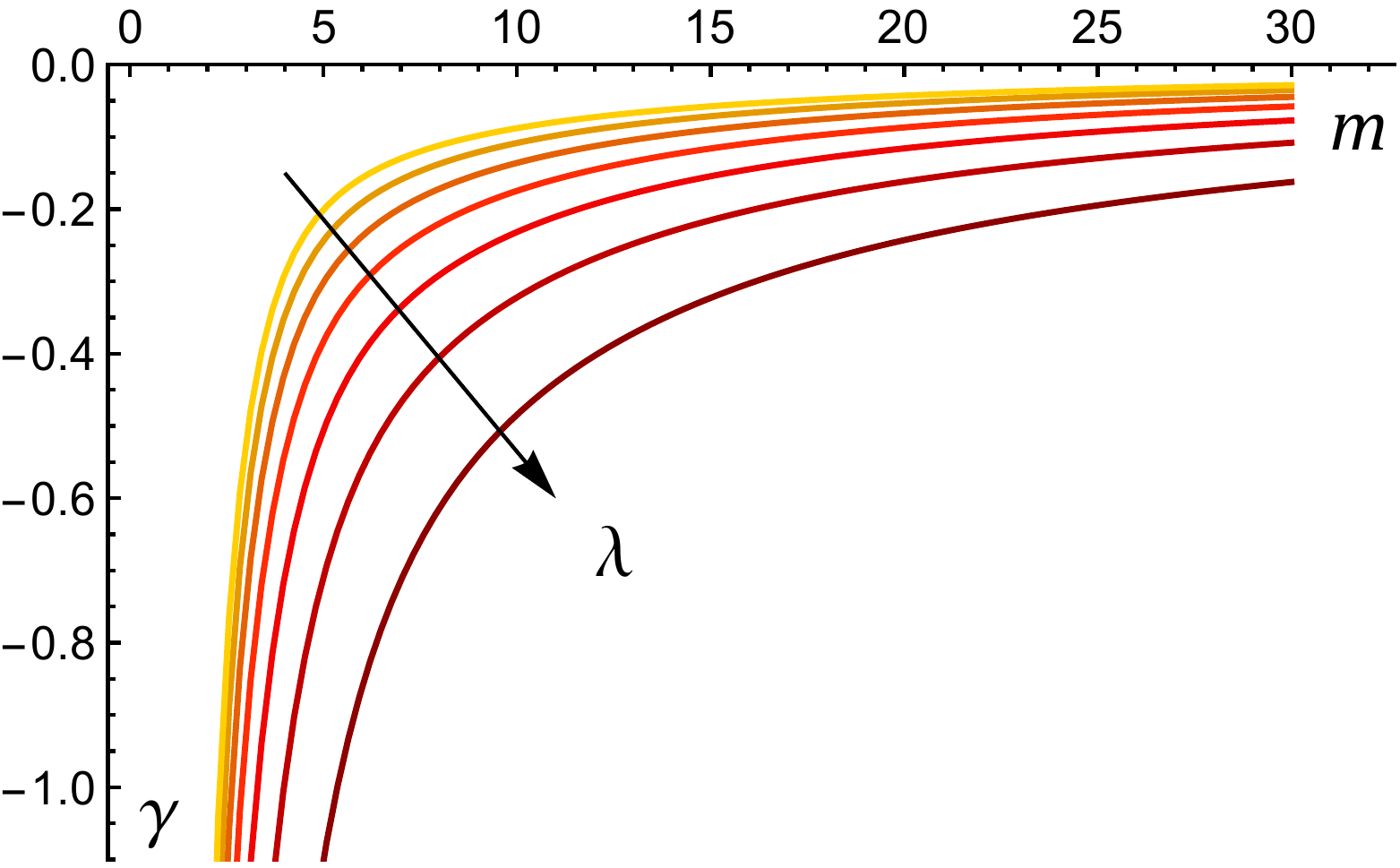}
\subcaption*{\textbf{(a)}}
\end{subfigure}\hfill
\begin{subfigure}[t]{0.47\textwidth}
\includegraphics[width=\linewidth, valign=t]{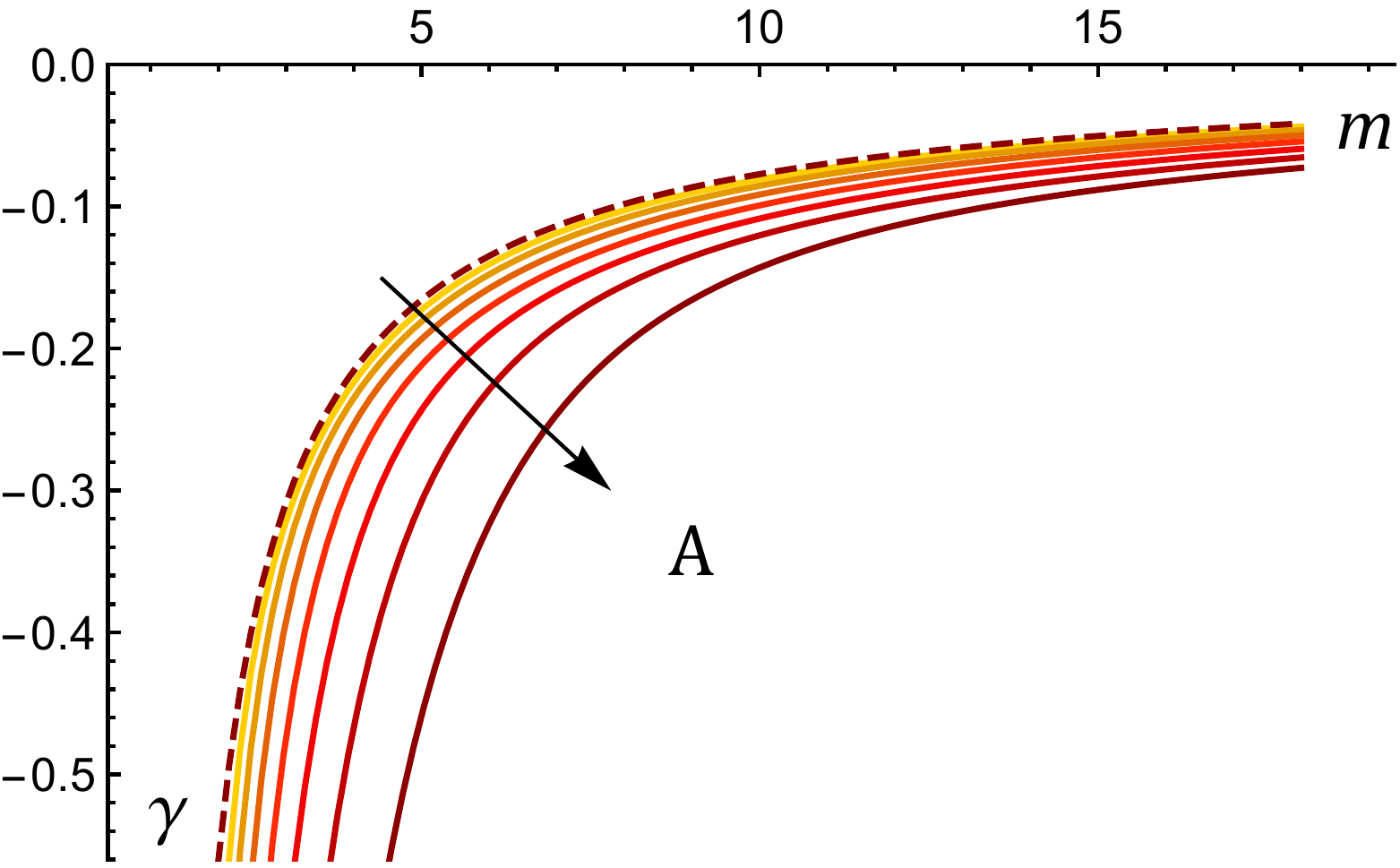}
\vspace{0.25mm}
\subcaption*{\textbf{(b)}}
\end{subfigure}
\caption{The variation of $\gamma$ with respect to $m$. In \textbf{(a)} we set $A=0.6$ and increase $\lambda$ from $0.5$ to $1.7$ in increments of $0.2$. In \textbf{(b)} we set $\lambda =1.5$ and increase $A$ from $0.2$ to $0.8$ in increments of $0.1$. The left most dashed curved corresponds to $A=0$, which is the special case of a solid cylinder. Arrows indicate the direction of parameter growth.}
\label{fig3}
\end{figure}

In Figure $\ref{fig1}$ \textbf{(a)} we plot $\gamma$ against $m$ for various fixed $\lambda\geq 1$ and $A=0.3$. It is observed that circumferential mode solutions are entirely possible, and the critical mode number $m_{cr}=2$ consistently. This corresponds to the tubes cross section bifurcating into an elliptic shape. Axial stretching is shown to have a destabilising influence on the tube in that $\gamma_{cr}$ decreases with increasing $\lambda$. In Figure $\ref{fig1}$ \textbf{(b)} we analyse the bifurcation behaviour over various tube thickness's. The variation of $\gamma_{cr}$ with respect to $A$ is found to be non-monotonic, with each curve having a maximum typically in the regime of moderate thickness. Thus, for a specific fixed $\lambda$, one can design a tube of such a thickness in order to provide added resistance against circumferential buckling.
 \\
\indent For completeness, we investigate whether circumferential mode solutions are also dominant for other tensile loading paths. To this end, we instead subject the tube first to a fixed surface tension $\gamma$ and zero axial force, and then increase the nominal axial stress $S_{zz}=S_{zz}(\lambda,\,a)$ as given by $(\ref{Nax})$ from zero. We note that in Figure $4$ \textbf{(b)} of \cite{EmeryFu2020}, the existence of \textit{axial} periodic modes with finite, non-zero wavenumber was found to be limited to the compressive regime under this loading path. Thus, we again state that if circumferential modes exist here in the tensile regime, they are dominant by default. In Figure $\ref{fig2}$ \textbf{(a)} we consider the variation of $S_{zz}$ with respect to $m$ for several fixed $\gamma$ and $A=0.5$. We observe that, for sufficiently small fixed $\gamma$, bifurcation into a circumferential mode solution with $m_{cr}=2$ is triggered at some critical axial stress $S_{zz}^{cr}>0$. Indeed, $S_{zz}^{cr}$ is seen to decrease as $\gamma$ increases, thus surface tension has a destabilising effect on the tube in this sense. At some fixed surface tension threshold, the tube becomes highly unstable and bifurcation into the elliptic mode occurs at $S_{zz}^{cr}=0$. This is shown by the lower-most dashed curve in \textbf{(a)} which corresponds to $\gamma=0.142375$. In Figure $\ref{fig2}$ \textbf{(b)}, the critical stress for bifurcation into the mode $m=2$ is seen to be non-monotonic with respect to $A$, with each curve having a maximum typically in the range of moderate thickness. For each fixed $\gamma$ under consideration, there exists a critical value of $A$ above which the elliptic mode cannot occur when the tube is in tension. 
\textbf{\subsection{ Case 2: Radially fixed inner boundary free of surface tension}}
The situation in case $2$ is found to be far different to that displayed previously for case $1$. To facilitate the analysis here, we take $\gamma$ as the load parameter, and in Figure $\ref{fig3}$ we plot $\gamma$ against $m$ for several tube thickness's and fixed axial stretches. We see that, across the wide range of parameter values considered, bifurcation into a circumferential mode solution is associated with negative critical surface tension values, which is physically implausible. For completeness, we include in Figure $\ref{fig3}$ \textbf{(b)} the special case of a solid cylinder corresponding to $A=0$ (dashed curve). Thus, we conclude that circumferential mode solutions are not possible in case 2, and that localisation in the tensile regime is consequently preferred. 
\textbf{\subsection{ Case 3: Radially fixed outer boundary free of surface tension}}

\subsection*{\textit{Taking $S_{zz}$ as the load parameter}}
\begin{figure}[h!]
\centering
\begin{subfigure}[t]{0.46\textwidth}
\includegraphics[width=\linewidth, valign=t]{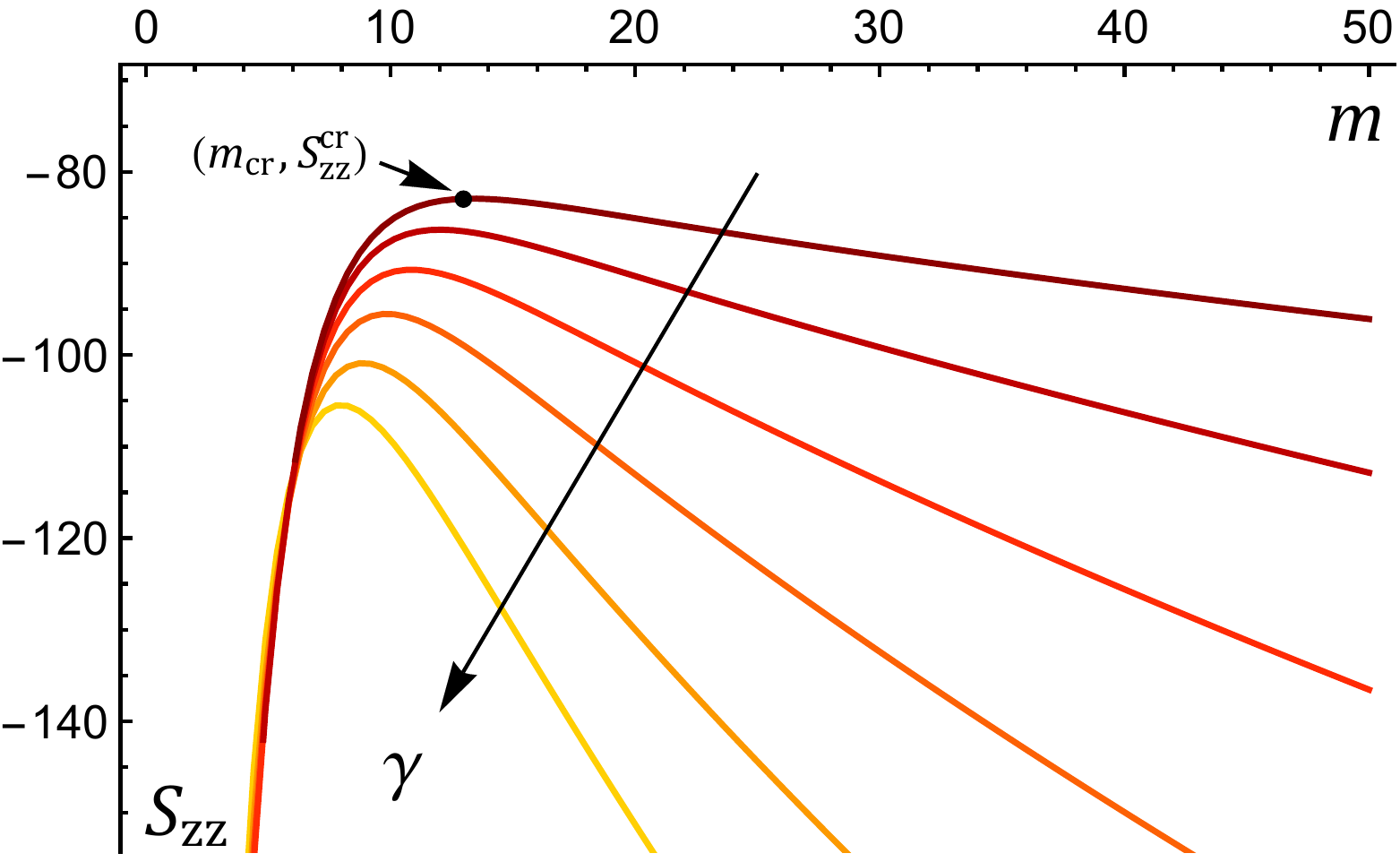}
\vspace{1mm}
\subcaption*{\textbf{(a)}}
\end{subfigure}\hfill
\begin{subfigure}[t]{0.455\textwidth}
\includegraphics[width=\linewidth, valign=t]{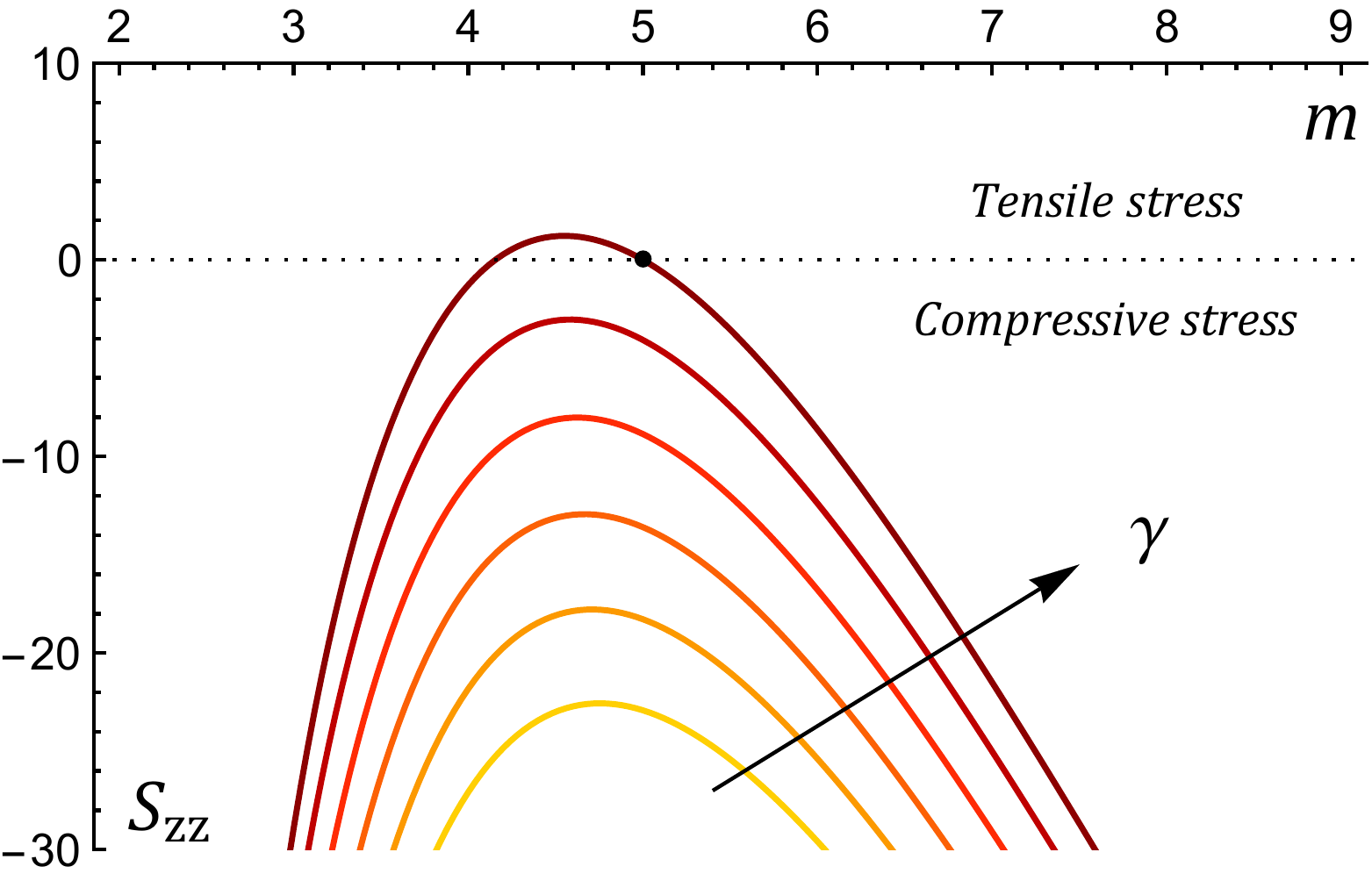}
\subcaption*{\textbf{(b)}}
\end{subfigure}
\caption{The variation of $S_{zz}$ with respect to $m$ for $A=0.8$ and \textbf{(a)} $\gamma=0.02,\,0.05,\,0.1,\,0.175,\,0.3,\,0.5$, \textbf{(b)} $\gamma=3.5,\,3.6,\,3.7,\,3.8,\,3.9,\,3.9844$. In both plots, the arrow indicates the direction of parameter growth. In \textbf{(a)}, the dot marks the critical pair $\left(m_{cr},\,S_{zz}^{cr}\right)=\left(13,\,-82.9374\right)$ for $\gamma=0.02$. The dot in \textbf{(b)} shows that, at the fixed surface tension threshold $\gamma=3.9844$, bifurcation into a circumferential mode with $m_{cr}=5$ is triggered solely by surface tension.}
\label{fig4}
\end{figure}

\begin{figure}[h!]
\centering
\begin{subfigure}[t]{0.45\textwidth}
\includegraphics[width=\linewidth, valign=t]{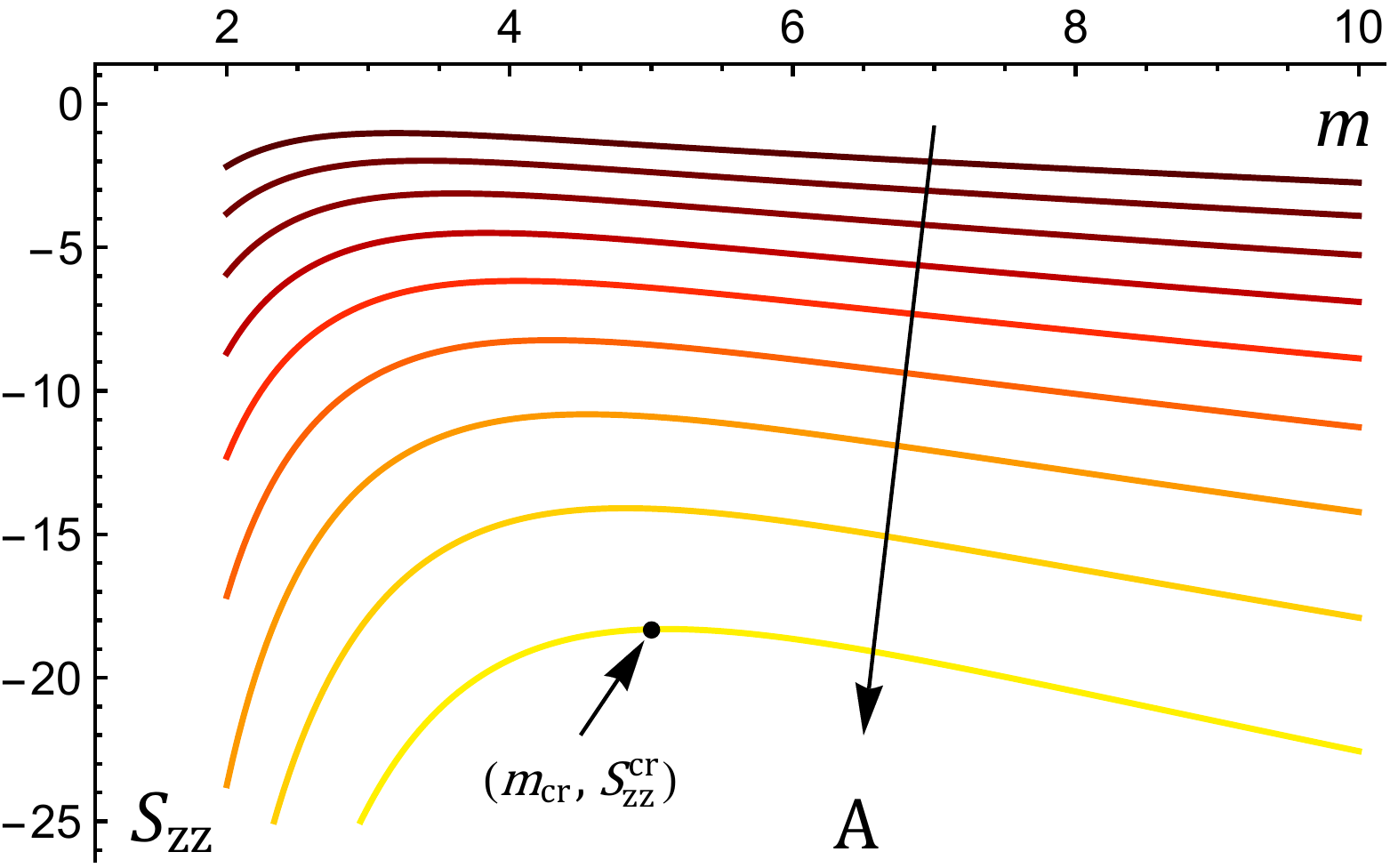}
\subcaption*{\textbf{(a)}}
\vspace{2mm}
\end{subfigure}\hfill
\begin{subfigure}[t]{0.445\textwidth}
\includegraphics[width=\linewidth, valign=t]{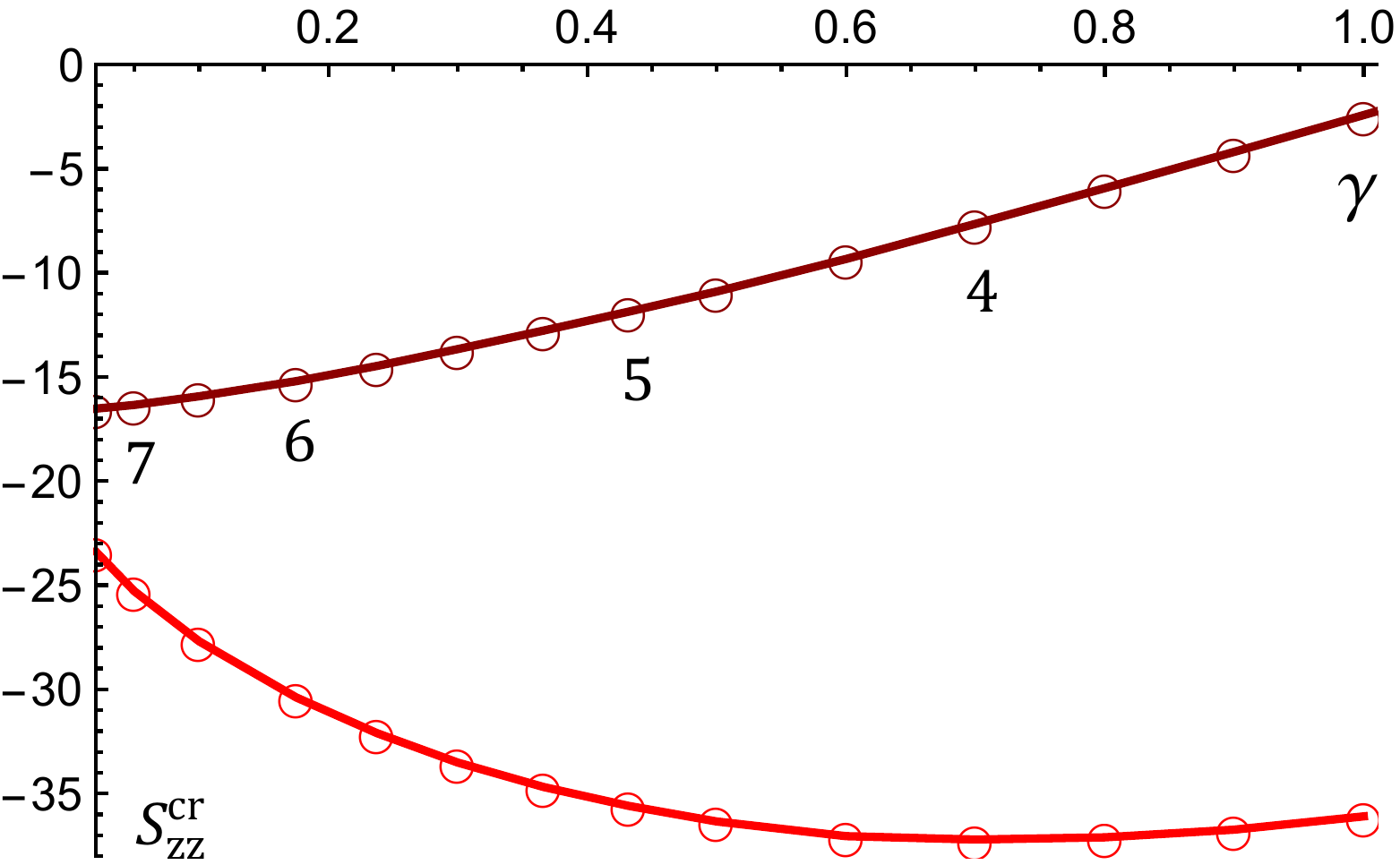}
\subcaption*{\textbf{(b)}}
\vspace{2mm}
\end{subfigure}
\vfill
\begin{subfigure}[t]{0.445\textwidth}
\includegraphics[width=\linewidth, valign=t]{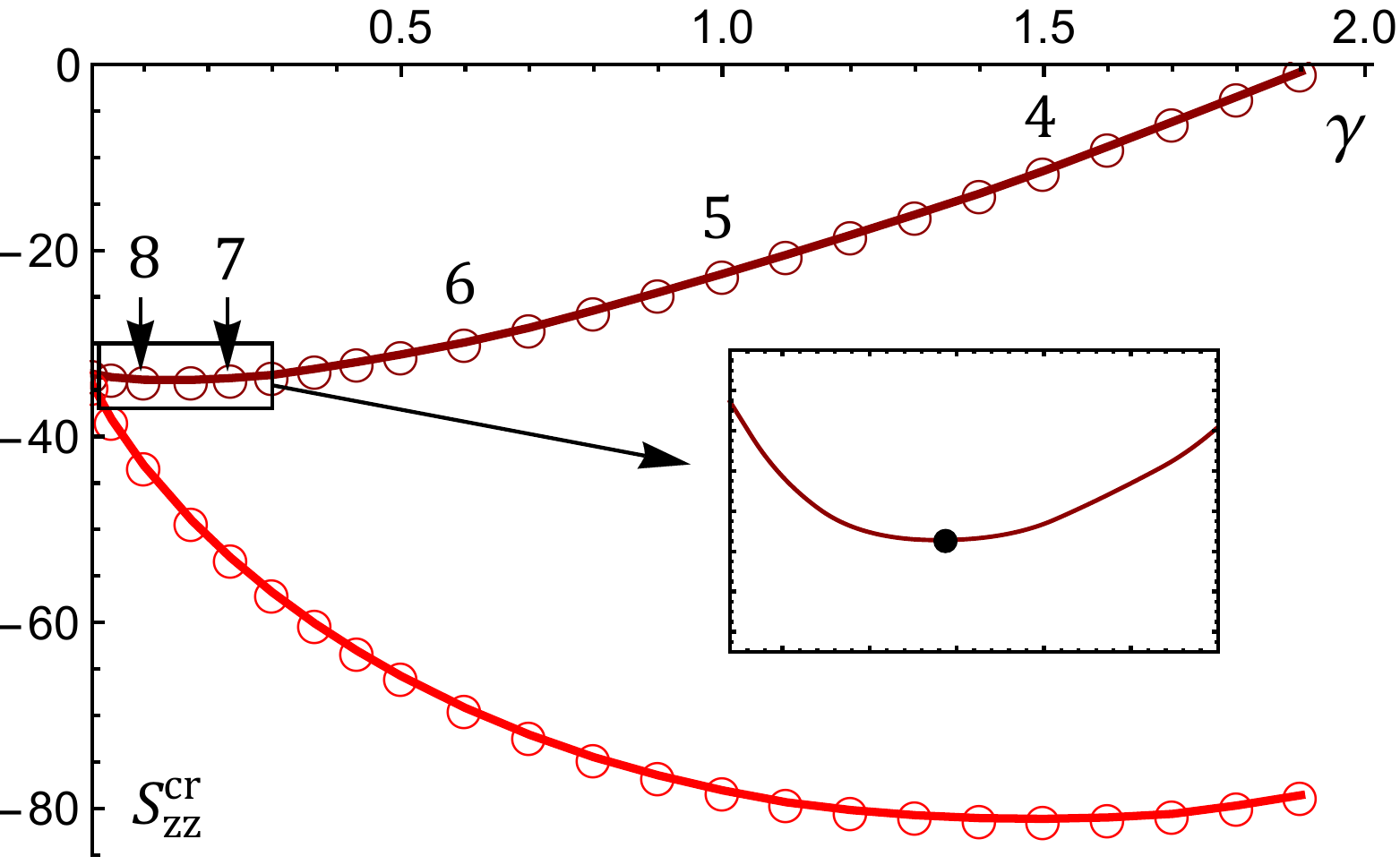}
\vspace{1mm}
\subcaption*{\textbf{(c)}}
\end{subfigure}\hfill
\begin{subfigure}[t]{0.45\textwidth}
\includegraphics[width=\linewidth, valign=t]{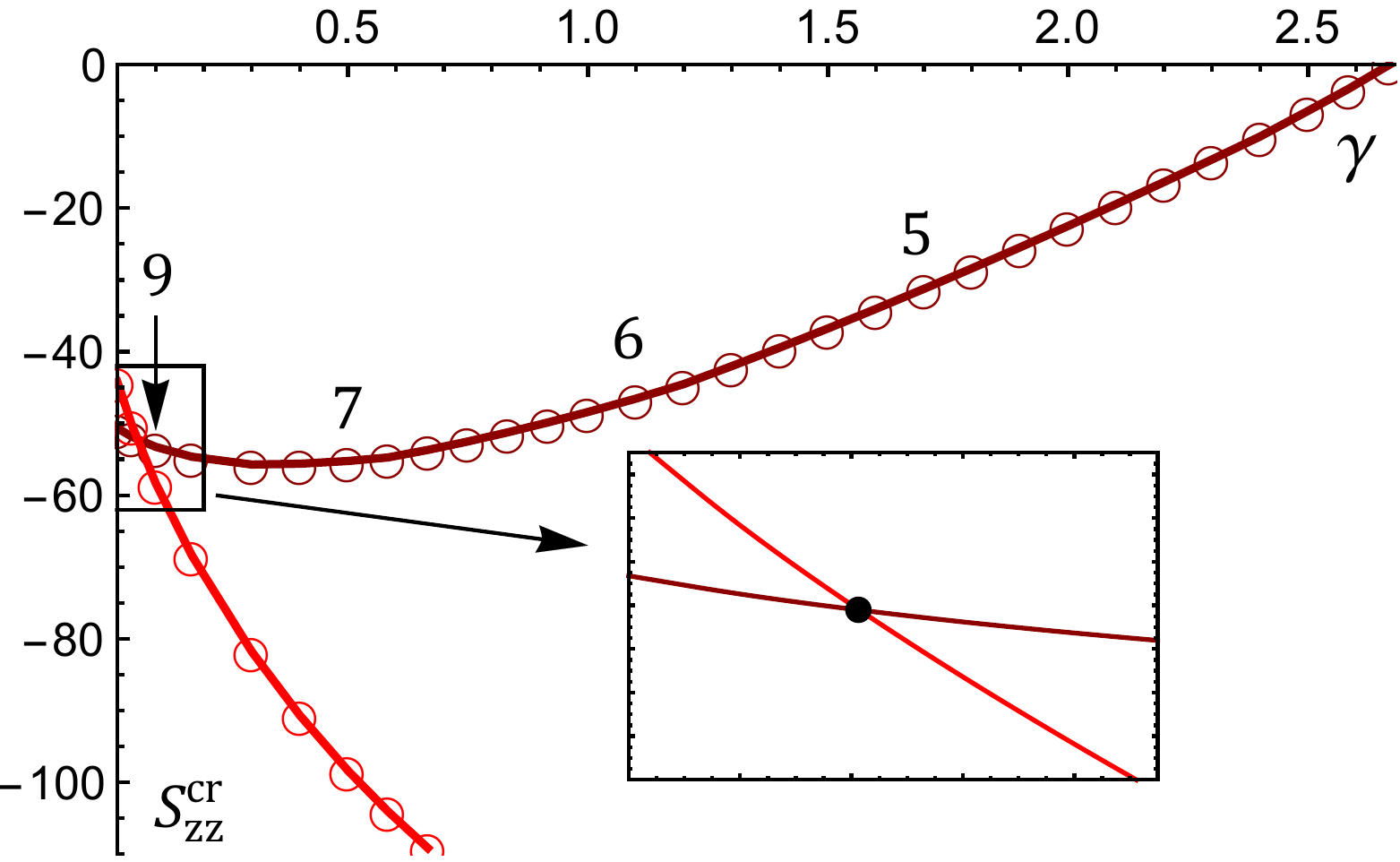}
\vspace{0.5mm}
\subcaption*{\textbf{(d)}}
\end{subfigure}
\caption{\textbf{(a)} The variation of $S_{zz}$ with respect $m$ for $\gamma=0.5$. The inner referential radius takes the fixed values $A=0.45,\,0.475,\,0.5,\,0.525,\,0.55,\,0.575,\,0.6,\,0.625,\,0.65$. The dot indicates the critical pair $(m_{cr},\,S_{zz}^{cr})=(5,\,-18.3092)$ for $A=0.65$ and the arrow indicates the direction of parameter growth. In the latter $3$ plots we show conditions for bifurcation into circumferential (maroon) and axial (red) modes, with the latter having critical wave numbers $k_{cr}\geq 0$. Specifically, we present the variation of $S_{zz}^{cr}$ with respect to $\gamma\geq 0.02$ for \textbf{(b)} $A=0.6$, \textbf{(c)} $A=0.7$ and \textbf{(d)} $A=0.75$. The inset in \textbf{(c)} shows that, below some critical tube thickness, this variation for circumferential buckling solutions is non-monotonic. The curve has a minimum at $\left(\gamma,\,S_{zz}^{cr}\right)=\left(0.143722,\,-33.95\right)$ as indicated by the black dot. The inset in \textbf{(d)} shows that, for a sufficiently thin tube, there exists a fixed surface tension threshold $\gamma=\gamma_1$ below which axial modes dominate circumferential buckling. Said threshold is the intersection between the two curves as indicated by the black dot, and corresponds to $\gamma_1=0.061327$ in this case. The integers labelled along each orange curve give the value of $m_{cr}$ at the corresponding critical load. The critical circumferential mode number is seen to decrease as the fixed surface tension gets larger.}
\label{fig5}
\end{figure}

We start by considering a representative case $A=0.8$, and show in Figure $\ref{fig4}$  the variation of $S_{zz}$ with respect to $m$ for several fixed $\gamma$.
We observe from Figure $\ref{fig4}$ \textbf{(a)} that, for considerably small fixed $\gamma$, bifurcation into a circumferential mode with finite wave number and $S^{cr}_{zz}<0$ is \textit{theoretically} possible. That is, bifurcation into a finite circumferential mode can occur at some critical \textit{compressive} axial stress. In Figure $\ref{fig4}$ \textbf{(b)} we see that, for sufficiently large fixed surface tension, circumferential buckling can exist in the tensile regime. However, at the fixed surface tension threshold $\gamma=3.9844$ the tube becomes highly unstable and bifurcation into a finite circumferential mode with $m_{cr}=5$ is triggered before an axial stress can be applied. \\
\indent In Figure $\ref{fig5}$ we analyse the bifurcation behaviour across various tube thickness's as well as the competition between circumferential and axial modes for $S_{zz}\leq0$. In Figure $\ref{fig5}$ \textbf{(a)} we plot $S_{zz}$ against $m$ for several fixed $A$ and $\gamma=0.5$. We observe that $S_{zz}^{cr}$ is a decreasing function of $A$, and thus the axial compressive stress required to trigger bifurcation into a finite circumferential mode is much smaller in thicker tubes. In Figures $\ref{fig5}$ \textbf{(b)}, \textbf{(c)} and \textbf{(d)} we compare the conditions for bifurcation into circumferential buckling modes (maroon) and axial modes with non-negative critical wave number (red) for three separate tube thickness's. As was observed in Figure \ref{fig4}, for each tube thickness considered there exists a fixed surface tension threshold at which the tube becomes highly unstable towards circumferential buckling, with $S_{zz}^{cr}=0$. Through inspection of \textbf{(b)} through to \textbf{(d)}, this threshold is seen to be a decreasing function of the tube thickness, echoing the findings in Figure \ref{fig5} \textbf{(a)} that thicker tubes are more susceptible to circumferential modes. We find also that, if the tube thickness is sufficiently small, the relationship between $S_{zz}^{cr}$ and $\gamma$ for circumferential modes is non-monotonic. This can clearly be seen by comparing Figures \ref{fig5} \textbf{(b)} and \textbf{(c)}, say, where $A=0.6$ and $0.7$ respectively. In the former, the magnitude of $S_{zz}^{cr}$ is a decreasing function of $\gamma$, whereas in the latter we see from the inset that $S_{zz}^{cr}$ has a minimum at $\left(\gamma,\,S_{zz}^{cr}\right)=\left(0.143722,\,-33.95\right)$. Thus, for $\gamma<0.143722$ (resp. $\gamma>0.143722$), larger fixed surface tension increases (resp. decreases) the axial load required for bifurcation. We observe that, for a sufficiently thick tube, circumferential buckling is triggered at lower compressions than localised beading or periodic axial modes for any fixed surface tension. However, for a sufficiently thin tube, there exists an interval $\gamma\in\left(0,\,\gamma_{1}\right)$ wherein axial modes dominate circumferential buckling. This can be clearly seen in Figure $\ref{fig5}$ \textbf{(d)} for $A=0.75$, where the inset shows an intersection between the two curves at $\gamma=\gamma_{1}=0.061327$. For $\gamma<\gamma_{1}$, the magnitude of $S_{zz}^{cr}$ is evidently less for axial modes than circumferential buckling.
 \\
\indent In the tensile regime, we comment that the fixed surface tension threshold at which circumferential modes are triggered at $S_{zz}^{cr}=0$ is typically less than the minimum $\gamma$ required for localisation. For instance, for $A=0.7$ it is shown in \cite{EmeryFu2020} that localisation is globally absent for $\gamma<6.66298$, but from Figure \ref{fig5} \textbf{(c)} we see that circumferential buckling is triggered purely by surface tension at $\gamma\approx 1.9$. Thus, bifurcation into circumferential mode solutions will have already occurred before the tensile stress required to trigger localisation can be applied, and so the former is dominant in this sense.
\subsection*{\textit{Taking $\gamma$ as the load parameter}}
\begin{figure}[h!]
\centering
\begin{subfigure}[t]{0.445\textwidth}
\includegraphics[width=\linewidth, valign=t]{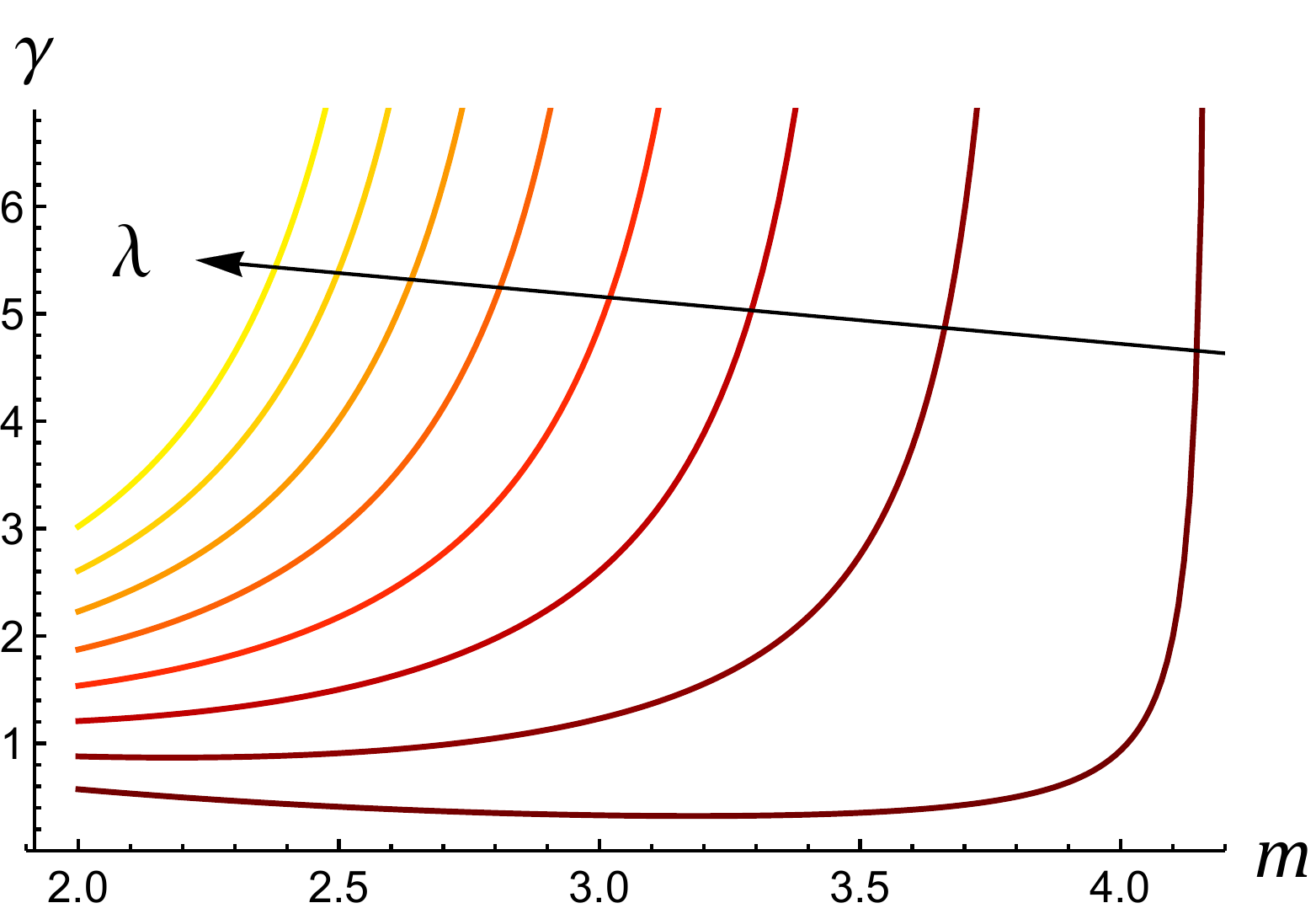}
\subcaption*{\textbf{(a)}}
\end{subfigure}\hfill 
\begin{subfigure}[t]{0.46\textwidth}
\includegraphics[width=\linewidth, valign=t]{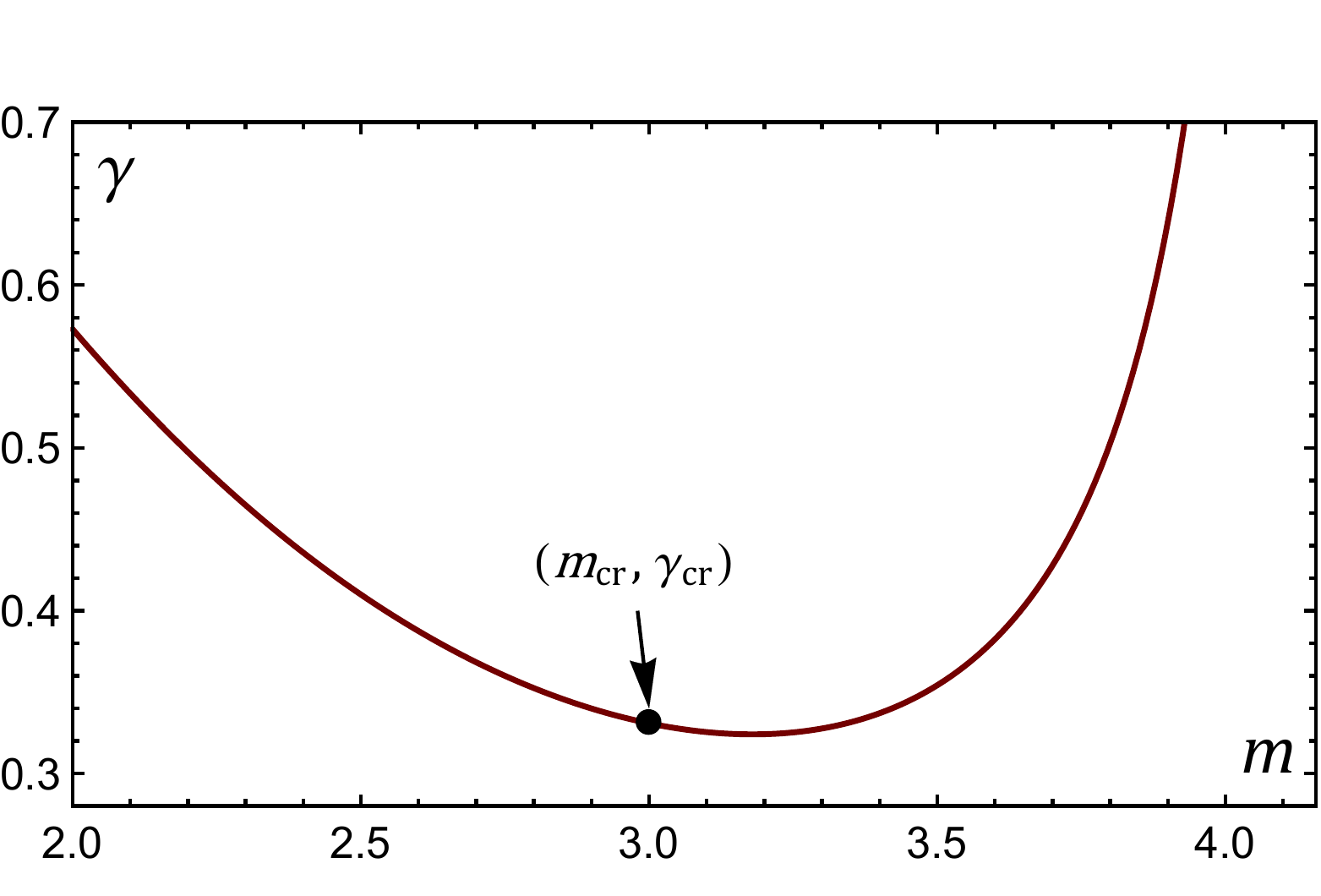}
\subcaption*{\textbf{(b)}}
\end{subfigure}
\begin{subfigure}[t]{0.445\textwidth}
\includegraphics[width=\linewidth, valign=t]{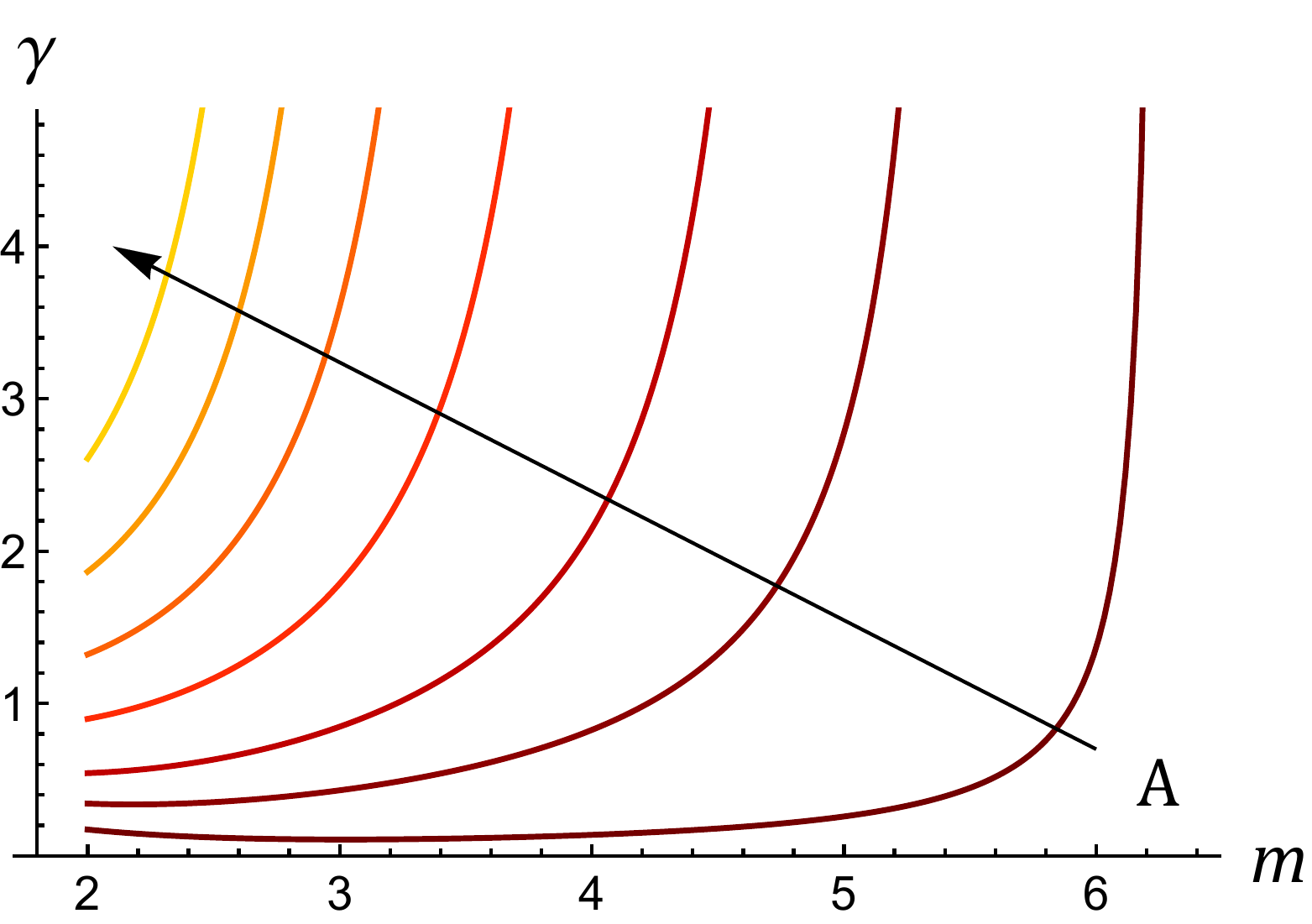}
\subcaption*{\textbf{(c)}}
\end{subfigure}\hfill 
\begin{subfigure}[t]{0.47\textwidth}
\includegraphics[width=\linewidth, valign=t]{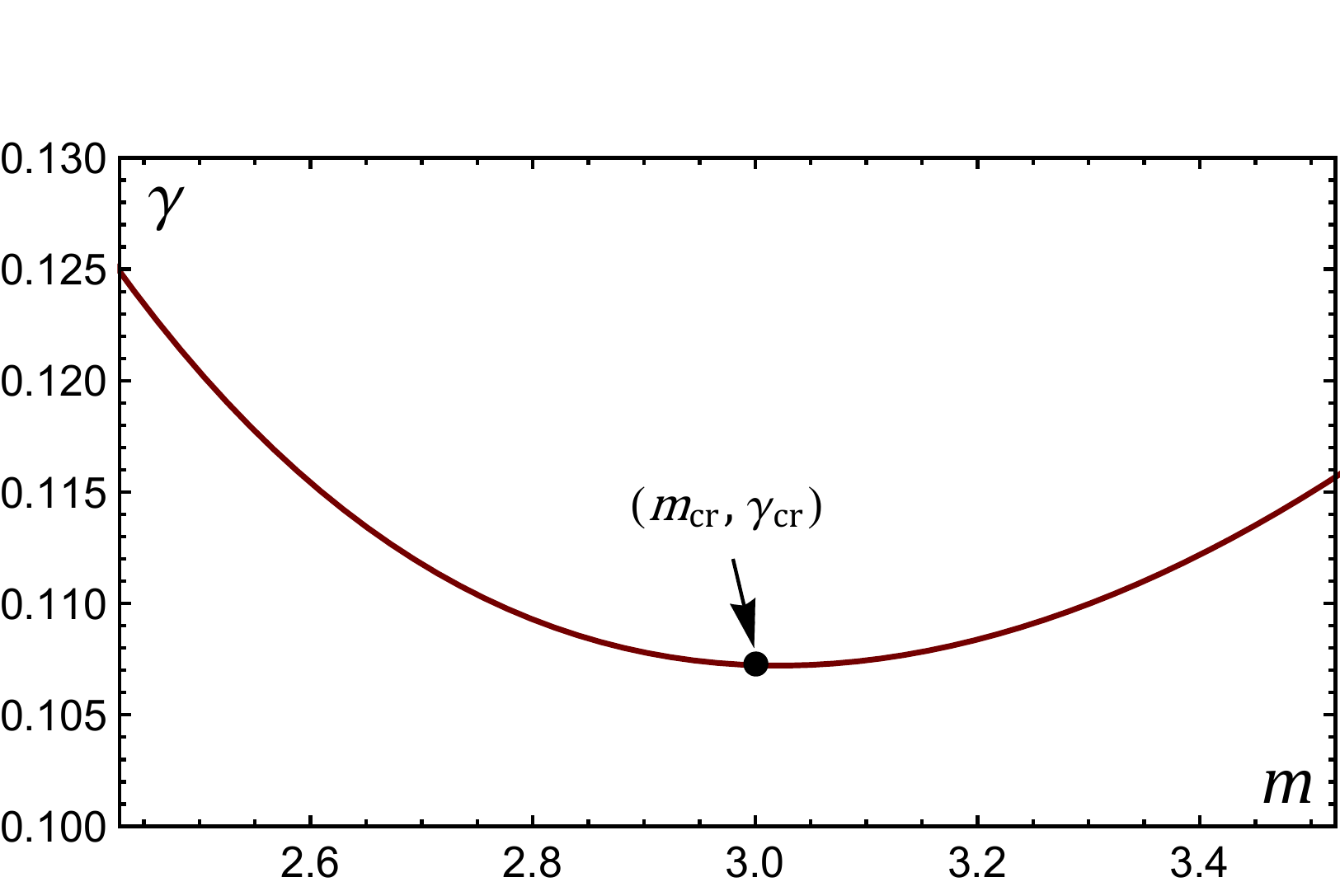}
\subcaption*{\textbf{(d)}}
\end{subfigure}
\caption{\textbf{(a)} The variation of $\gamma$ with respect $m$ for $A=0.4$. The axial stretch $\lambda$ takes the fixed values $\lambda=0.891,\,0.9,\,0.905,\,0.91,\,0.92,\,0.93,\,0.94,\,0.95\,\,\text{and}\,\,0.96$. \textbf{(b)} A blow up of the curve in \textbf{(a)} corresponding to $\lambda=0.891$ about the critical pair $\left(m_{cr},\,\gamma_{cr}\right)=\left(3,\,0.330843\right)$. \textbf{(c)} The variation of $\gamma$ with respect to $m$ for $\lambda=0.95$ and $A=0.2725,\,0.285,\,0.3,\,0.325,\,0.35,\,0.375\,\,\text{and}\,\,0.4$. \textbf{(d)} A blow up of the curve in \textbf{(c)} corresponding to $A=0.2725$ about the critical pair $\left(m_{cr},\,\gamma_{cr}\right)=\left(3,\,0.107227\right)$.}
\label{fig6}
\end{figure}

\begin{figure}[h!]
\centering
\begin{subfigure}[t]{0.455\textwidth}
\includegraphics[width=\linewidth, valign=t]{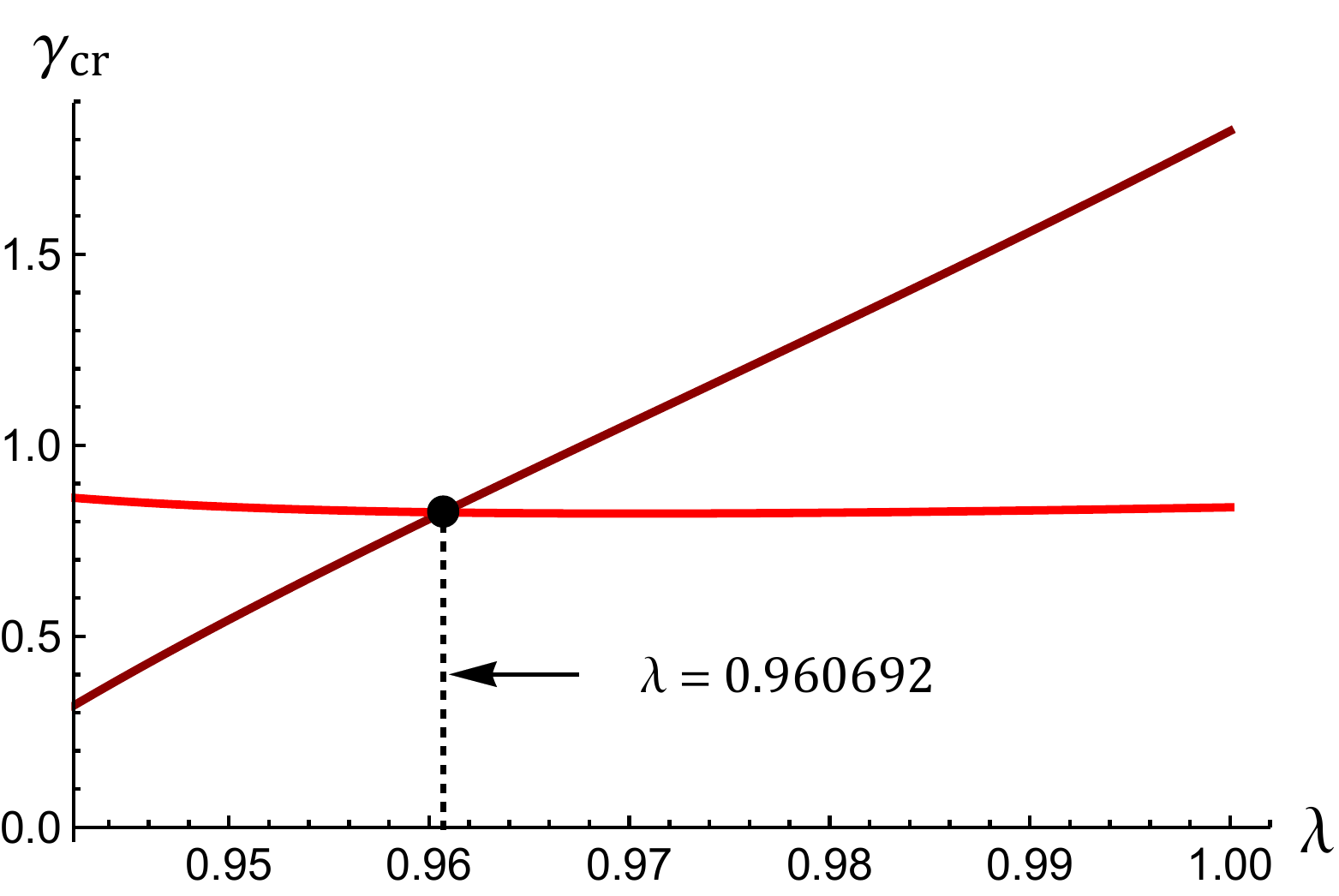}
\subcaption*{\textbf{(a)}}
\end{subfigure}\hfill
\begin{subfigure}[t]{0.455\textwidth}
\includegraphics[width=\linewidth, valign=t]{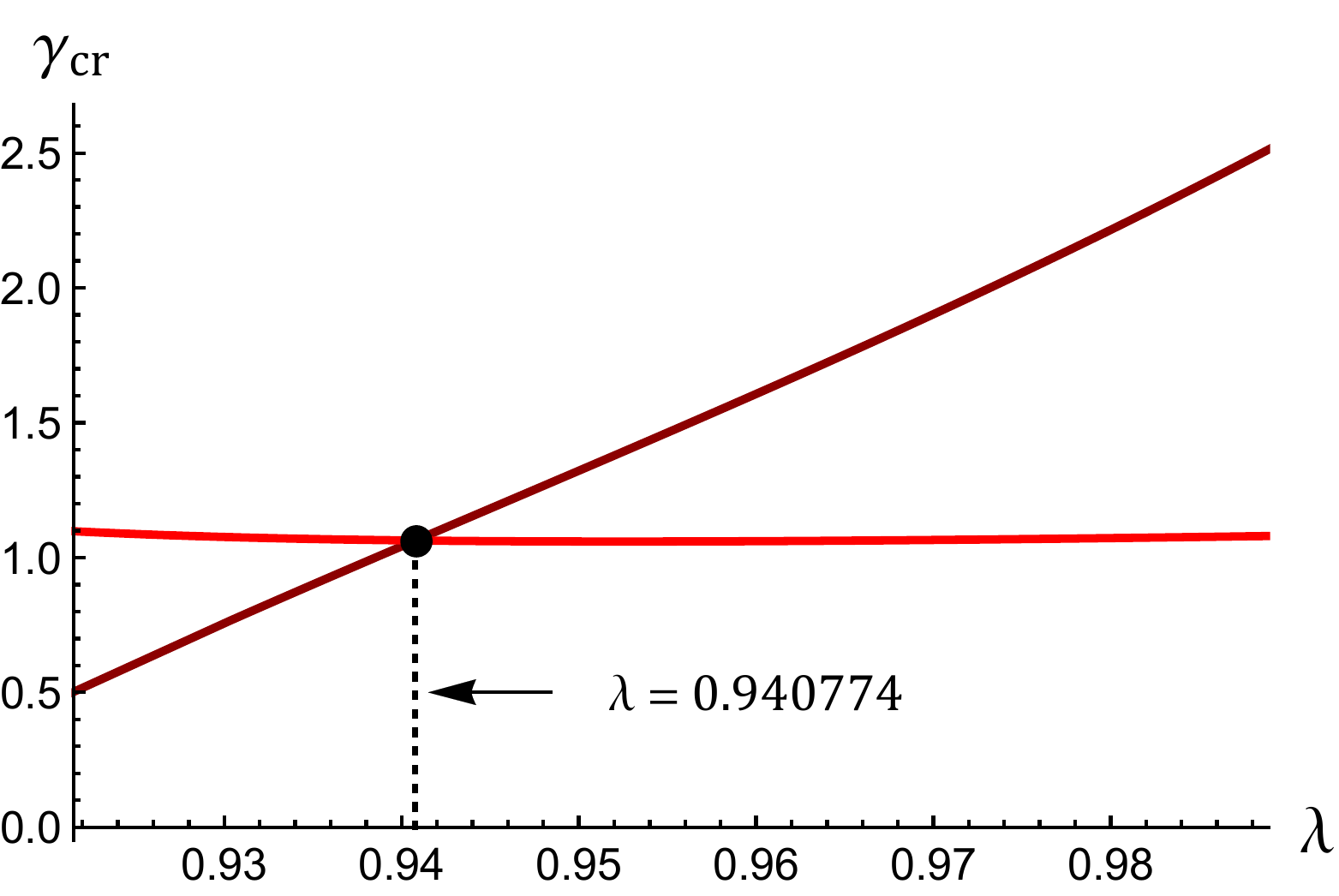}
\subcaption*{\textbf{(b)}}
\end{subfigure}
\begin{subfigure}[t]{0.455\textwidth}
\includegraphics[width=\linewidth, valign=t]{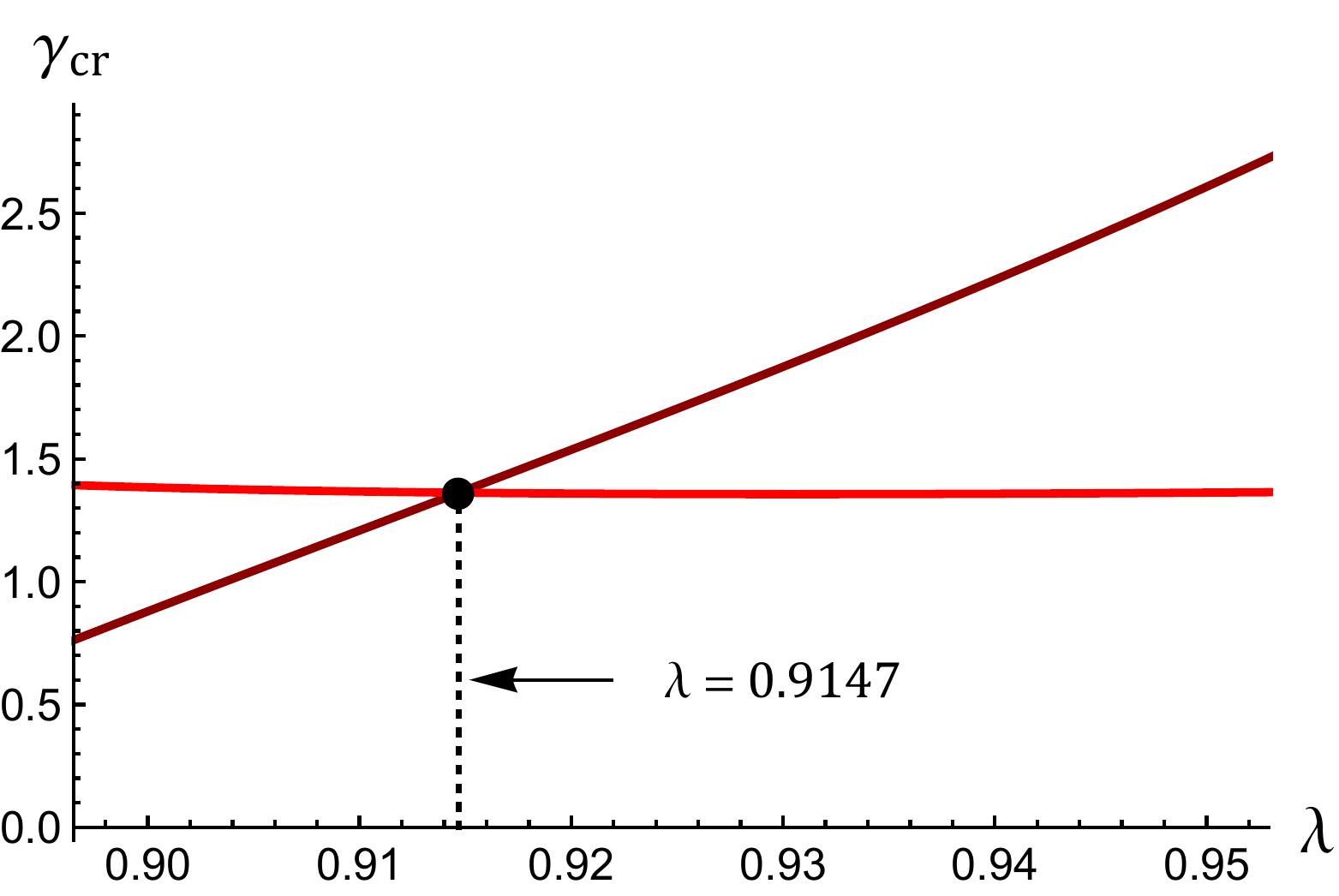}
\subcaption*{\textbf{(c)}}
\end{subfigure}\hfill
\begin{subfigure}[t]{0.455\textwidth}
\includegraphics[width=\linewidth, valign=t]{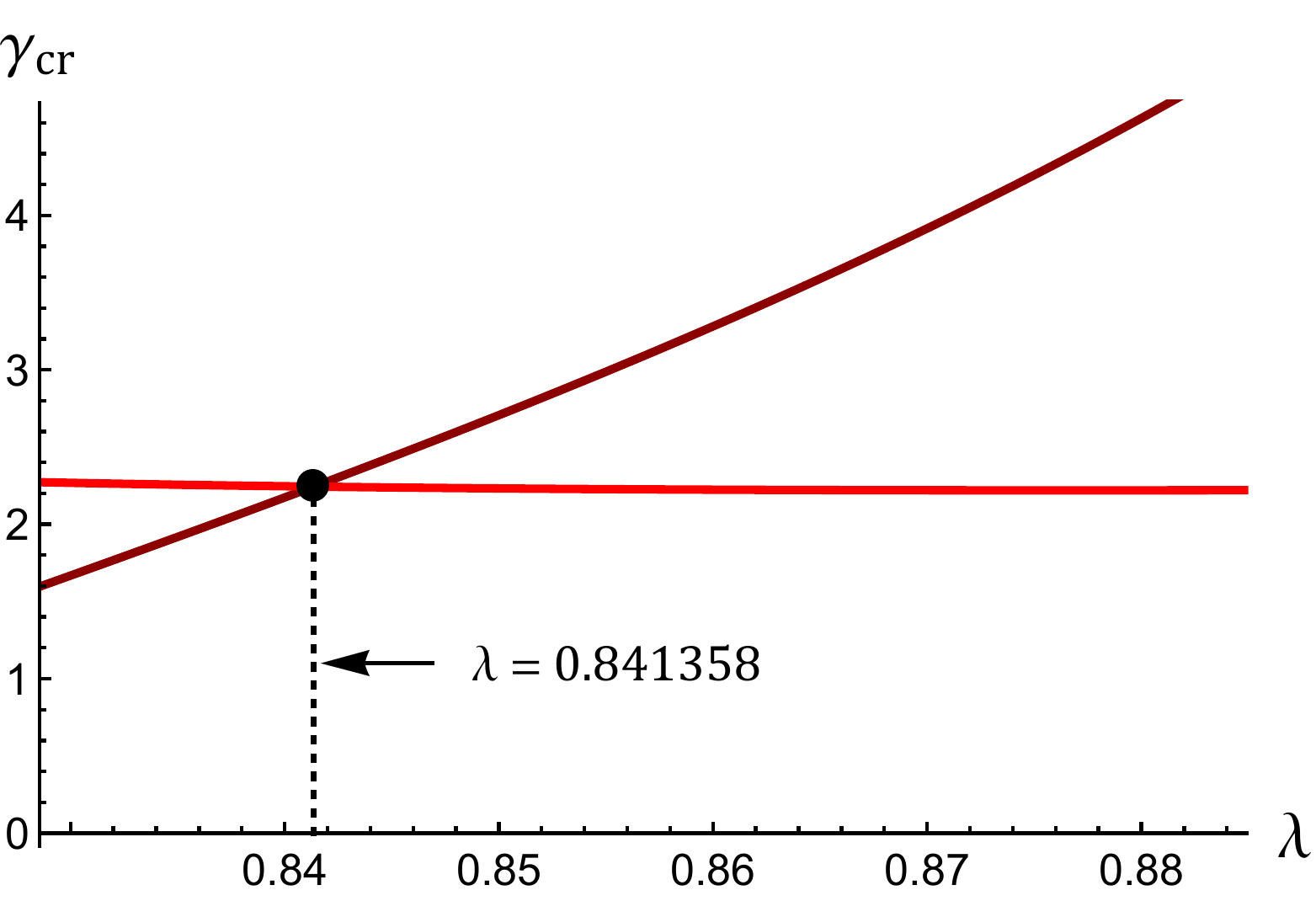}
\subcaption*{\textbf{(d)}}
\end{subfigure}
\caption{Conditions for bifurcation into localised and periodic axial modes (red) and the elliptic circumferential buckling mode (maroon) in the $(\lambda,\,\gamma_{cr})$ plane for \textbf{(a)} $A=0.3$, \textbf{(b)} $A=0.35$, \textbf{(c)} $A=0.4$ and \textbf{(d)} $A=0.5$. }
\label{fig7}
\end{figure}

Alternatively, we may subject the tube to a fixed axial stretch $\lambda$ and then gradually increase the surface tension from zero. We consider first a representative case $A=0.4$, and in Figure $\ref{fig6}$ \textbf{(a)} we plot $\gamma$ against $m$ for several fixed $\lambda<1$. We observe that, for a sufficiently small fixed compression, the critical circumferential mode number $m_{cr}=2$. For larger compressive stresses, we may have $m_{cr}>2$, and this is demonstrated in Figure $\ref{fig6}$ \textbf{(b)} where we show that $m_{cr}=3$ for $\lambda=0.891$. If we compress even further, the tube enters a highly unstable regime whereby $\gamma_{cr}=0$.
In Figure $\ref{fig6}$ \textbf{(c)} we analyse the bifurcation behaviour across various tube thickness's. For the fixed stretch $\lambda=0.95$, we observe that $m_{cr}=2$ consistently below some critical tube thickness. Above this threshold, we may have $m_{cr}>2$, and this is observed in Figure $\ref{fig6}$ \textbf{(d)} where we show that $m_{cr}=3$ for $A=0.2725$. We also deduce that thicker tubes are more sensitive towards circumferential mode solutions; for a sufficiently thick tube, circumferential buckling is triggered at $\gamma_{cr}=0$. \\
\indent It thus remains to assess the competition between axial and circumferential mode solution to deduce absolutely the preference of the tube in this loading scenario. In Figure $\ref{fig7}$ we plot the critical surface tension for bifurcation into axial modes with non-negative wave number (red) and circumferential modes (maroon) against $\lambda$ for several tube thickness's. For each fixed value of $A$ considered, we find that there exists a fixed compression threshold beyond which circumferential mode solutions are favoured over localisation or periodic axial modes. Prior to this point, localisation is preferred over circumferential and periodic axial modes for all tube thickness's.
\section{Conclusions}
The objective of this study was two-fold. Firstly, for the three distinct boundary conditions under consideration, we endeavoured to determine whether bifurcation of soft slender tubes into finite circumferential buckling modes can necessarily occur for multiple loading paths. Secondly, where circumferential mode solutions do exist, an analysis of their competition with localisation and periodic axial modes was desired.\\
\indent In case 1 we focussed on the regime of tensile axial loads given the preference of slender tubes towards Euler buckling over periodic wrinkling when in compression. We determined that the tube was highly susceptible to circumferential mode solutions when either the surface tension $\gamma$ or the nominal axial stress $S_{zz}$ was taken as the load parameter. Specifically, the elliptic mode $m_{cr}=2$ was found to be unanimously favoured. In either of these loading scenarios, the fixed axial stretch $\lambda$ and surface tension $\gamma$ (respectively) were found to have a destabilising effect on the tube. Where $S_{zz}$ is taken as the load parameter, we determined that the tube becomes highly unstable to the elliptic mode at a fixed surface tension threshold, and bifurcation is triggered prior to the application of an axial stress at this point. For the representative case $A=0.5$, this threshold was determined to be $\gamma=0.142375$. With the aid of results given in \cite{EmeryFu2020}, we determined absolutely that for both loading scenarios circumferential buckling dominates localised and periodic axial modes.\\
\indent In case 2 we took $\gamma$ as the load parameter, and found over a wide range of tube thickness's and fixed axial stretches that bifurcation into circumferential modes was associated with negative values of $\gamma_{cr}$. Given the physical implausibility of negative surface tension, we deduced from this that such a bifurcation cannot necessarily take place in case 2. This conclusion was also shown to be valid for the special case of a solid cylinder on taking the limit $A\rightarrow 0$.  \\
\indent In case 3 we extended our interests to the compressive stress regime given the known competition between circumferential and axial wrinkling modes when the tube is purely under axial loading \cite{liu2018axial}. We chose first $S_{zz}$ as the load parameter; it was shown that, for sufficiently small fixed $\gamma$, circumferential buckling modes may emerge at some critical \textit{compressive} axial stress $S_{zz}^{cr}<0$. At some fixed surface tension threshold, the tube becomes highly unstable with $S_{zz}^{cr}=0$; this threshold was deduced to be $\gamma=0.39844$ for the representative case $A=0.8$, with the preferred mode number being $m_{cr}=5$. Thicker tubes were also ascertained to be more sensitive towards such solutions. In this loading scenario, circumferential mode solutions were found to be consistently dominant over periodic axial modes and localisation in the compressive regime provided that the tube is thick enough. However, for sufficiently thin tubes, there exists a fixed surface tension interval $\gamma\in\left(0,\,\gamma_{1}\right)$ wherein axial modes instead dominate circumferential buckling. Where $\gamma$ is instead taken as the load parameter, we determined that there exists a critical fixed axial compression below which the critical buckling mode is $m_{cr}=2$. Unsurprisingly, bifurcation became more likely as said compression increased, and for a small enough value of $\lambda$ we deduced that $\gamma_{cr}=0$. Similar behaviour was deduced across different tube thickness's. We finally showed that, for several tube thickness's, there exists a critical fixed compression beyond which circumferential modes are dominant over localisation and periodic axial modes.

\begin{acknowledgements}
The first author (DE) acknowledges the School of Computing and Mathematics, Keele University for supporting his PhD studies through a faculty scholarship.
\end{acknowledgements}

%
%

\bibliographystyle{spmpsci}      
\bibliography{refer}   

%
%

\end{document}